\documentclass[fleqn,usenatbib]{mnras}

\usepackage[T2A]{fontenc}
\usepackage[utf8]{inputenc}
\usepackage[english]{babel}
\usepackage{hyperref}

\usepackage{graphicx}	% Including figure files
\usepackage{subfigure}
\usepackage{amsmath}	% Advanced maths commands
\usepackage{amssymb}	% Extra maths symbols
\hypersetup{draft}
\usepackage{multirow}

\usepackage{xcolor}
\definecolor{mygray}{gray}{0.6}
\definecolor{myblue}{RGB}{20, 20, 200}
\definecolor{myred}{RGB}{200, 20, 20}

\title[New cLBVs in NGC\,4449]{Search for LBVs in the Local Volume galaxies: study of four stars in NGC 4449}

\author[Y. Solovyeva et al.]{
Y. Solovyeva,$^{1}$\thanks{E-mail:solovyeva@sao.ru}
A. Vinokurov,$^{1}$
A. Sarkisyan,$^{1}$
A. Kostenkov,$^{1,2}$
K. Atapin,$^{3}$
S. Fabrika $^{1,4}$ 
\newauthor
D. Oparin,$^{1}$
A. Valeev,$^{1,5}$
D. Bizyaev,$^{3,6}$
P. Nedialkov$^{7}$
and O. Spiridonova$^{1}$
\\
$^{1}$Special Astrophysical Observatory, Nizhnij Arkhyz, Russia\\
$^{2}$Mathematics\&Mechanics Faculty, Saint Petersburg State University, 7/9 Universitetskaya Emb., Saint Petersburg 199034, Russia\\
$^{3}$Sternberg Astronomical Institute, Lomonosov Moscow State University, Universitetskij Pr. 13, Moscow 119992, Russia\\
$^{4}$Institute of Physics, Kazan Federal University, Kremlevskaya 18, 420008 Kazan, Russia\\
$^{5}$Crimean Astrophysical Observatory, Russian Academy of Sciences, Nauchnyi 298409, Russia\\
$^{6}$Apache Point Observatory and New Mexico State University, Sunspot, NM 88349, USA\\
$^{7}$Department of astronomy, Sofia University, Sofia 1164, Bulgaria\\}

\date{Accepted XXX. Received YYY; in original form ZZZ}

\pubyear{2021}

\begin{document}
\label{firstpage}
\pagerange{\pageref{firstpage}--\pageref{lastpage}}
\maketitle

\begin{abstract}

We continue to search for LBV stars in galaxies outside the Local Group. In this work, we have investigated four luminous stars in NGC\,4449. Multiple spectral observations carried out for J122810.94+440540.6, J122811.70+440550.9, and J122809.72+440514.8 revealed the emission features in their spectra that are characteristic of LBVs. Photometry showed noticeable brightness changes of J122809.72+440514.8 ($\Delta I=0.69\pm0.13^m$) and J122817.83+440630.8 ($\Delta R=2.15\pm0.13^m$), while the variability of J122810.94+440540.6 and J122811.70+440550.9 does not exceed $0.3^m$ regardless of the filter. We have obtained estimates of the interstellar reddening, photosphere temperatures, and bolometric luminosities $\log(\text{L}_\text{Bol}/\text{L}_{\odot}) \approx 5.24-6.42$. Using the CMFGEN code, we have modelled the spectrum of the cold state of J122809.72+440514.8 ($T_{\text{eff}}=9300\,$K) and have obtained  possible value of the mass loss rate $\dot{M} = 5.2\times10^{-3}\,M_{\odot}\,yr^{-1}$. Based on the observational properties, J122809.72+440514.8 and J122817.83+440630.8 were classified as LBVs, while the other two stars were classified as LBV candidates or B[e]-supergiants candidates.

\end{abstract}


%%%%%%%%%%%%%%%%%%%%%%%%%%%%%%%%%%%%%%%%%%%%%%%%%%

%%%%%%%%%%%%%%%%% BODY OF PAPER %%%%%%%%%%%%%%%%%%

\section{Introduction}

Luminous blue variables are the massive ($M\geq25M\odot$, \citealt{Humphreys16}) stars with high luminosity ($\gtrsim10^5 L_{\odot}$), characterized by significant spectral and photometric variability.

A characteristic feature of many LBV stars is the S Dor-type variability \citep{vanGenderen01}. During this cycle, both photometric and spectral variability are observed. The brightness variation amplitude in the V band can reach 2.5 magnitudes, with the cold state of the star corresponding to maximum brightness, and the hotter state to minimum. Spectra of LBVs are similar to those of the A-F supergiants, hot B supergiants, or Of/late--WN stars depending on the photosphere temperature \citep{Vink12,Humphreys94}. A luminous star exhibiting the described variability can be unambiguously classified as an LBV in the S Dor cycle. In addition, some LBV stars experience dramatic brightness changes in the form of giant eruptions of more than $2.5^m$. This type of brightness change is $\eta$-Car variability \citep{Humphreys99}. An important feature of the spectral energy distribution of LBVs is the absence of an IR excess associated with the emission of hot dust \citep{Humphreys14}.

The evolutionary status of LBVs remains poorly known. In the accepted view they correspond to the transitional phase from single massive O-stars to the Wolf-Rayet stars\citep{Groh14}. Some studies have shown that rotating LBVs with initial masses of 20-25 M$_{\odot}$ can evolve directly into the core-collapse supernovae bypassing the Wolf-Rayet stage \citep{Groh13}.The possibility of the appearance of LBV as a result of the evolution of close binaries is considered in \citet{Smith15}.

The classification of high luminosity stars observed in the galaxies M\,33 and M\,31 was proposed in \citet{Humphreys14}, according to spectral and photometric characteristics of stars: B[e]-supergiants, Of/late--WN, LBV stars, warm hypergiants, Fe\,II-emission stars, hot and intermediate supergiants. These stars have similar spectral characteristics, which complicates the search for LBVs; however, only LBVs have significant brightness variability. Despite their similarity, the evolutionary connections between these luminous stars have not yet been clarified.

To date, only about 40 LBVs and about a hundred LBV candidates (cLBVs)  are known in our and other galaxies, mostly belonging to the Local Group \citep{Richardson18}. The information about these stars in the Local Volume are incomplete, and only a few LBVs and cLBVs are known beyond 1 Mpc (for example, \citealt{Pustilnik17, Humphreys19, Drissen97, Goranskij16, Solovyeva19}). The confirmation of the LBV status requires a large amount of observational time for revealing photometric and spectral variability, but the discovery of new LBVs and candidates may clarify the origin of the LBV phenomenon and evolutionary status of LBVs.

\begin{figure*} 
\center{\includegraphics[angle=0, width=0.7\linewidth]{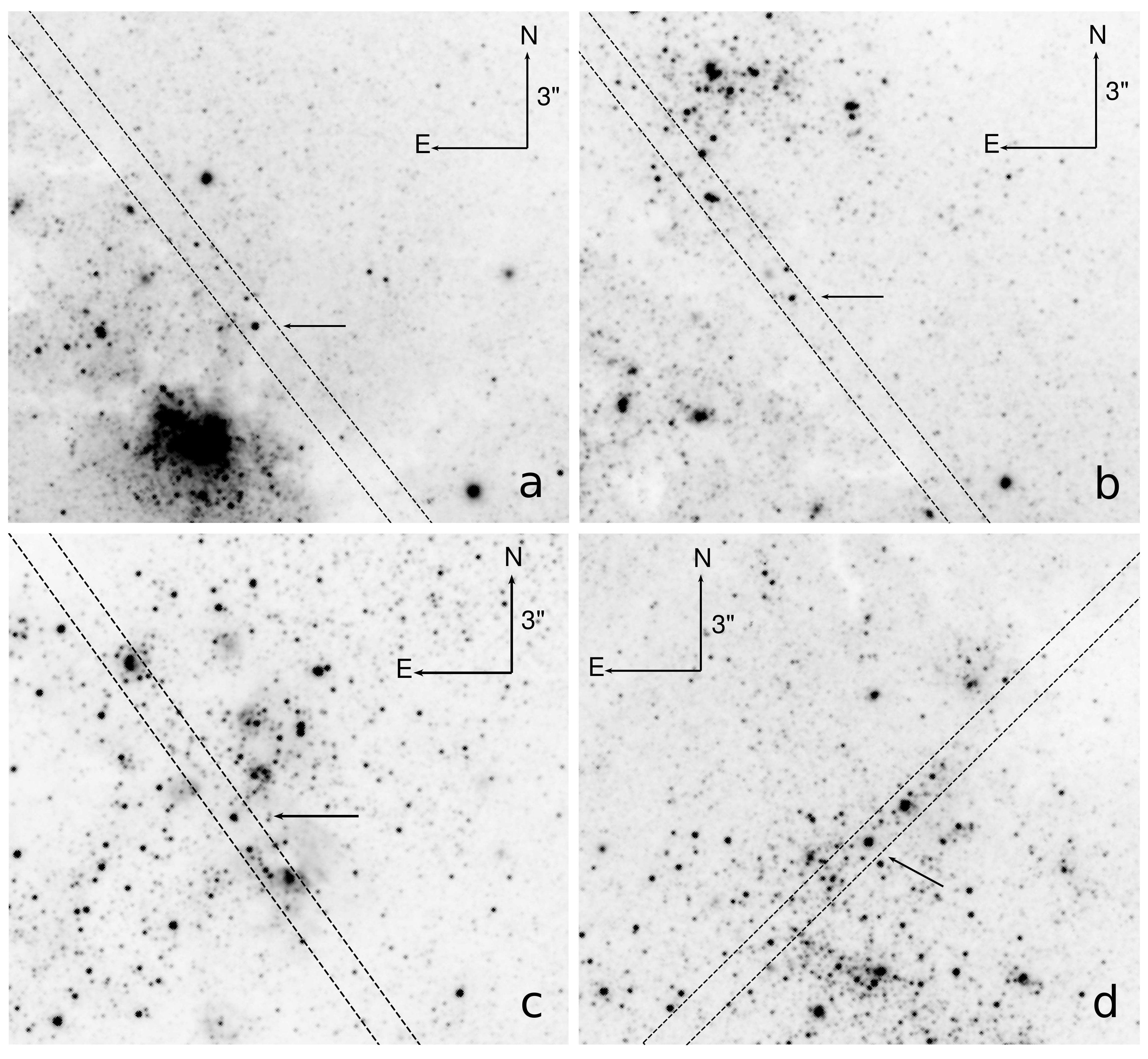}}
\caption{Images in \textit{HST/F550M}: J122810.94+440540.6 (a), J122811.70+440550.9 (b), J122817.83+440630.8 (c), J122809.72+440514.8 (d). The 1.0 arcsec
slit is shown.} 
\label{Fig1} 
\end{figure*}

We search for LBVs and similar objects in the Local Volume galaxies by  selecting the point-like H$\alpha$ sources associated with blue stars. For our purposes we use archival broadband and narrowband H$\alpha$ images obtained with the Hubble Space Telescope (HST). The first results of our search in the NGC\,4736 and NGC\,247 galaxies were published in the works \citet{Solovyeva19, Solovyeva20}. This paper presents the results of a study of the NGC\,4449 galaxy (distance D=4.27 Mpc, \citet{Tully13}). This dwarf irregular galaxy of Magellanic-type (Ibm type) has specific star formation rate higher than that of the Large Magellanic Cloud (according to Catalog \& Atlas of the LV galaxies\footnote{https://serv.sao.ru/lv/lvgdb/}). We have discovered 4 new cLBVs in this galaxy: J122810.94+440540.6, J122811.70+440550.9, J122809.72+440514.8 and J122817.83+440630.8 (Fig. \ref{Fig1}). In this paper, we present the results of a detailed study of the detected stars based on spectral and photometric observations.

%Fig.1

\section{OBSERVATIONS AND DATA REDUCTION}
\subsection{Spectroscopy}

The spectra of the all four LBV candidates were obtained with the 6-m telescope of SAO RAS (BTA) using the SCORPIO or SCORPIO-2 focal reducers \citep{Afanasiev05,AfanasievSco2}. The slit size was 0.5--1.2\arcsec. The observation dates, seeing and used grisms are shown in Table \ref{Tab1}. Spectral data processing was performed with the context \textsc{long} of \textsc{midas} using the standard algorithm. The spectra were extracted using the \textsc{spextra} package \citep{Sarkisyan17} intended to deal with long-slit spectra in the crowded stellar fields.

\subsection{Imaging}

Photometric data were obtained with the 2.5-m telescope of Caucasian Mountain Obseravtory of SAI MSU (2.5-m CMO), the BTA and the Zeiss-1000 of SAO RAS (Zeiss-1000). We also used archival data of HST (ACS, WFPC2 and WFC3 cameras) and the Bok Telescope of Kitt Peak National Observatory (Bok).  Details are given in Tables~\ref{Tab2}, \ref{Tab3} and \ref{Tab4}. Primary processing of the 2.5-m CMO, BTA and Zeiss-1000 data was carried out with \textsc{midas}.

To determine stellar magnitudes in the ground-based observations, we performed point spread function (PSF) photometry since the objects are located in crowded stellar fields, and aperture photometry leads to overestimation of the source fluxes. The PSF photometry was performed using the \textsc{daophot\,ii} package \citep{Stetson87}. Absolute calibration for the ground-base observations were performed using 27--32 reference stars whose fluxes were measured from the HST images.  The only exception was the U-band data (observation taken with 2.5-m CMO, Table~\ref{Tab3}) where we used reference star fluxes from the SDSS Photometric Catalog \citep{Abazajian09} in order to avoid incorrect flux estimation for stars with deep Balmer jump caused by the fact that the U filter is centered ($\lambda_{c}\approx3600$\AA) near the Balmer jump while the HST F336W (or F330W) filter cover these wavelength only partly. The resulting magnitudes are shown in Table~\ref{Tab3} with errors include statistical errors of the flux measurements, errors of absolute calibration and also account for background irregularities around the objects. For the objects J122810.94+440540.6 and J122811.70+440550.9 we did not provide result of ground-based photometry because they have nearby stars of comparable brightness and many faint stars, which complicates the background subtraction and makes results unstable. For these sources we show only magnitudes obtained from the HST data.

For the HST/WFPC2 data, we performed PSF photometry using the \textsc{hstphot}\,1.1 package \citep{Dolphin2000} on c0f images. The data from the ACS and WFC3 cameras were analyzed by the aperture photometry method using the APPHOT package of \textsc{iraf}. The calibrated pipeline-processed drizzled images were acquired from the MAST archive. Radii of circular apertures ($r_{\rm ap}$) for the source flux measurements and radii of annuli ($r_{\rm in}, r_{\rm out}$) for the background were chosen depending on the camera: $r_{\rm ap} = 0.15\arcsec$, 0.075\arcsec\ and 0.12\arcsec, $r_{\rm in} = 0.25\arcsec$, 0.25\arcsec\ and 0.20\arcsec, $r_{\rm out} = 0.45\arcsec$, 0.50\arcsec\ and 0.44\arcsec\ for ACS/WFC, ACS/HRC and WFC3/UVIS, respectively.
The aperture corrections were taken into account by photometric measurements of 20 (ACS/WFC), 35 (ACS/HRC), and 36 (WFC3/UVIS) single bright stars.

The object J122809.72+440514.8 has a saturated central pixel in the F435W, F555W, F814W images of the HST/ACS/WFC data obtained on 2005 November 11.
To exclude damaged part of its PSF, we carried out photometry of this source in the annular aperture with inner and outer radii of 0.1\arcsec\ and 0.2\arcsec. To obtain valid absolute magnitudes, this aperture was calibrated on $\sim40$ reference stars. The sky background for this source was measured from the annulus with $r_{\rm in}=0.3$\arcsec\ and $r_{\rm out}=0.5$\arcsec. Correctness of this approach is confirmed by the fact that after the conversion to the standard (Johnson-Cousins) system, the V magnitudes from the damaged observation (obtained from the F555W image) and from the closest subsequent observation (the F500M filter, November 18) appeared to be consistent within measurement errors.

The measured HST magnitudes were converted into the standard Johnson-Cousins system using the calcphot function of the \textsc{PySynphot} package. For J122810.94+440540.6 and J122817.83+440630.8, we assumed a power law as the model spectrum, and determined spectral indices from the fluxes in adjacent filters. Another two objects J122811.70+440550.9 and J122809.72+440514.8 (according to the 1997 data) show evidences of Balmer discontinuity (see below). Therefore, to convert the F336W or F330W magnitudes into the U band, we used the model spectra presented in Sec.~\ref{res}. The rest magnitudes were converted using a power law with spectral indices calculated in the manner described above.

Resulting magnitudes are shown in Tables~\ref{Tab3} and \ref{Tab4}. The provided errors include, beside the statistical error of the flux measurement, the accuracy of the conversion between the photometric systems \citep{Sirianni05, Harris18}, the stability of zero points\footnote{https://www.stsci.edu/hst/instrumentation/wfc3/data-analysis/photometric-calibration,\\ https://www.stsci.edu/itt/review/ihb\_cy15/
ACS/c03\_intro\_acs6.html} and the stability of the filter PSF in each particular observation. (for example, \citealt{Anderson06}). Thus, since the instrumental errors have a characteristic value of the order of 1-3\%, we obtained the total photometry errors not better than 3\% even in the cases where the statistical errors were insignificant.

Also, we analysed infrared images of three LBV candidates obtained with the WFC3/IR camera on 2019 April 4 (Table~\ref{Tab4}). J122817.83+440630.8 was outside the field of view of this pointing. We carried out aperture photometry using aperture radii $r_{\rm ap}=0.26\arcsec$, $r_{in}=0.43$\arcsec\ and $r_{in}=0.82$\arcsec. The aperture correction factor was determined by measurements of $\sim$ 20 single bright stars in 0.26\arcsec\ and 0.40\arcsec\ apertures. The obtained magnitudes are shown in Table~\ref{Tab4}. In this table we also present the results of photometry in the $H\alpha$ filter (F658N) of the ACS/WFC and ACS/HRC cameras.

\begin{table*}
\caption{Log of spectral observations on the BTA.}
\begin{tabular}{ccccccc}
\hline\hline 
 Star& Date& Spectrograph, grism & Spectral  & Spectral  & Seeing & Total exp., s\\  
  & &  & resolution,\AA & range, \AA  &  & \\  \hline
 J122810.94+440540.6 & 2014/01/02 & SCORPIO/VPHG1200R & 5.0 & 5700-7400 & 1.6\arcsec & 2400 \\
  & 2015/01/18 & SCORPIO/VPHG1200G & 5.0 & 3900-5700 & 1.7\arcsec & 2400 \\
  & 2020/01/18 & SCORPIO/VPHG1200B & 5.5 & 3600-5400 & 1.8\arcsec &  1800\\ \hline
  
 J122811.70+440550.9 & 2014/01/02 & SCORPIO/VPHG1200R & 5.0 & 5700-7400 & 1.6\arcsec & 2400\\
  & 2015/01/18 & SCORPIO/VPHG1200G & 5.0 & 3900-5700 & 1.7\arcsec & 2400\\
  & 2020/01/18 & SCORPIO/VPHG1200B & 5.5 & 3600-5400 & 1.8\arcsec & 1800\\ \hline 
 J122817.83+440630.8 & 2021/02/11 & SCORPIO-2/VPHG1200@540 & 5.2 & 3650-7250 & 2.3\arcsec & 1200\\  \hline
  
 J122809.72+440514.8 & 2014/01/02 & SCORPIO/VPHG1200R & 5.0 & 5700-7400 & 1.5\arcsec & 2400 \\
  & 2015/01/18 & SCORPIO/VPHG1200G & 5.0 & 3900-5700 & 1.7\arcsec & 2400 \\
  & 2017/03/31 & SCORPIO/VPHG1200G & 5.0 & 3900-5700 & 1.8\arcsec & 3000 \\ 
  & 2018/02/18 & SCORPIO/VPHG1200G & 5.0 & 3900-5700 & 1.1\arcsec & 2400\\  
  & 2018/02/18 & SCORPIO/VPHG550G & 7.3 & 3100-7300 & 1.2\arcsec & 1800\\ 
  & 2020/08/18 & SCORPIO/VPHG1200B & 5.5 & 3600-5400 & 1.6\arcsec & 2700\\ \hline  

\end{tabular}
\label{Tab1}
\end{table*}

\begin{table*}
\caption{Log of HST observations. Objects J122810.94+440540.6, J122811.70+440550.9, J122817.83+440630.8 and J122809.72+440514.8 are marked with 1, 2, 3 and 4.}
\begin{tabular}{cccc}
\hline\hline 
 Date& Camera& Filters& Objects\\   \hline
  1995/05/10 & WFPC2 & F606W & 1, 2, 3, 4 \\
  1997/07/28 & WFPC2 & F170W, F336W, F555W, F814W & 4\\
  1998/01/09 & WFPC2 & F170W, F336W, F555W, F814W & 1, 2, 3\\
  2005/11/10 & ACS/WFC & F435W, F555W, F814W & 1, 2\\
  2005/11/11 & ACS/WFC & F435W, F555W, F658N, F814W & 1, 2, 3, 4\\
  2005/11/17 & ACS/WFC & F814W & 1, 2, 3, 4\\
  2005/11/18 & ACS/WFC & F550M & 1, 2, 3, 4 \\
  2006/01/26 & ACS/HRC & F330W, F550M, F658N, F814W & 1, 2 \\ 
  2014/07/09 & WFC3/UVIS & F275W, F336W& 1, 2, 3, 4 \\
  2019/04/04 & WFC3/IR & F110W, F160W & 1, 2, 4\\
  \hline
\end{tabular}
\label{Tab2}
\end{table*}

\begin{table*}
\begin{minipage}{18cm}
\caption{Results of the optical and UV photometry. The columns show the instruments, dates and observed stellar magnitudes (not corrected for reddening). All magnitudes are given in the VEGAMAG system.}
\begin{tabular}{lcccccccc} \hline\hline 
 \centering
Telescope & Date  & F170W, mag & F275W, mag &  U, mag & B, mag & V, mag &R, mag & I, mag \\ \hline
\multicolumn{9}{c}{J122810.94+440540.6} \\ \hline

WFPC2 &1995/05/10 & ---& ---&---& --- &  --- & $19.15\pm0.04$ & --- \\
WFPC2 &1998/01/09 & $>20.00$& ---&$18.91\pm0.04$ & --- &  $19.34\pm0.03$ & --- & $18.60\pm0.03$ \\
ACS/WFC& 2005/11/10& ---& ---& --- & --- & $19.34\pm0.03$ & --- & $18.63\pm0.03$\\
ACS/WFC& 2005/11/11& ---& ---& --- & $19.67\pm0.03$ & $19.33\pm0.03$ & --- & $18.64\pm0.03$\\
ACS/WFC& 2005/11/17& ---& ---& --- & --- & --- & --- & $18.67\pm0.03$ \\
ACS/WFC& 2005/11/18& ---& ---& --- & --- & $19.45\pm0.03$ & --- & --- \\
ACS/HRC& 2006/01/26& ---& ---& $19.09\pm0.04$ & $19.60\pm0.03$ & $19.31\pm0.03$ & --- & $18.66\pm0.03$ \\
WFC3/UVIS & 2014/07/09& --- & $19.69\pm0.05$& $18.80\pm0.04$ & --- & ---  \\
\hline

\multicolumn{9}{c}{J122811.70+440550.9} \\ \hline

WFPC2 &1995/05/10 & ---& ---&---& --- &  --- & $19.74\pm0.04$ & --- \\
WFPC2 &1998/01/09 & $18.83\pm0.11$& ---&$19.19\pm0.04$ & --- &  $20.19\pm0.03$ & --- & $19.50\pm0.03$ \\
ACS/WFC& 2005/11/10& ---& ---& --- & --- & $20.12\pm0.03$ & --- & $19.48\pm0.03$\\
ACS/WFC& 2005/11/11& ---& ---& --- & $20.29\pm0.03$ & $20.15\pm0.03$ & --- & $19.51\pm0.03$\\
ACS/WFC& 2005/11/17& ---& ---& --- & --- & --- & --- & $19.47\pm0.03$ \\
ACS/WFC& 2005/11/18& ---& ---& --- & --- & $20.25\pm0.03$ & --- & --- \\
ACS/HRC& 2006/01/26& ---& ---& $19.33\pm0.04$ & $20.32\pm0.03$ & $20.28\pm0.04$ & --- & $19.60\pm0.03$ \\
WFC3/UVIS & 2014/07/09& --- & $18.40\pm0.05$& $19.12\pm0.04$ & --- & ---  \\
\hline

\multicolumn{9}{c}{J122817.83$+$440630.8} \\ \hline

WFPC2 &1995/05/10 & ---& ---&---& --- &  --- & $22.03\pm0.06$ & --- \\
WFPC2 &1998/01/09 & $>20.29$& ---&$21.25\pm0.08$ & --- &  $22.01\pm0.06$ & --- & $21.60\pm0.07$ \\
BOK &2001/03/31 &---&---&---&---&---&$19.88\pm0.12$& --- \\
ACS/WFC& 2005/11/11& ---& ---& --- & $20.98\pm0.03$ & $20.69\pm0.03$ & --- & $20.03\pm0.03$\\
ACS/WFC& 2005/11/17& ---& ---& --- & --- & --- & --- & $20.00\pm0.03$ \\
ACS/WFC& 2005/11/18& ---& ---& --- & --- & $20.72\pm0.03$ & --- & --- \\
WFC3/UVIS & 2014/07/09& --- & $21.28\pm0.05$& $21.79\pm0.03$ & --- & ---  \\
2.5m CMO & 2020/03/07 & --- & --- & ---& --- & ---& $22.04\pm0.20$&---\\
\hline

\multicolumn{9}{c}{J122809.72+440514.8} \\ \hline
WFPC2& 1995/05/10 & ---& ---& --- & --- & ---& $18.22\pm0.04$ & ---\\
WFPC2& 1997/07/28& $>20.20$& ---& $18.14\pm0.03$ & --- & $17.98\pm0.04$ & --- & $17.48\pm0.03$ \\
BOK &2001/03/31 &---&---&---&---&---&$17.71\pm0.11$& --- \\
ACS/WFC& 2005/11/11& ---& ---& --- & $18.30\pm0.08$ & $18.05\pm0.09$ & --- & $17.56\pm0.07$ \\
ACS/WFC& 2005/11/17& ---& ---& --- & --- & --- & --- & $17.50\pm0.03$ \\
ACS/WFC& 2005/11/18& ---& ---& --- & --- & $17.94\pm0.03$ & --- & --- \\
WFC3/UVIS & 2014/07/09& --- & $19.19\pm0.05$& $17.98\pm0.03$ & --- & ---  \\
BTA & 2018/02/18& --- & --- &--- & $18.61\pm0.15$ & $18.23\pm0.15$ & $18.01\pm0.11$ & $17.73\pm0.07$ \\
2.5m CMO & 2019/01/18 & --- & --- & ---&$18.49\pm0.11$ & $18.13\pm0.10$& $18.02\pm0.07$&---\\
Zeiss-1000 & 2019/04/09& --- & --- & --- &$18.53\pm0.09$ &$18.28\pm0.09$ & --- & --- \\
Zeiss-1000 & 2019/05/22& --- & --- & --- &$18.51\pm0.11$ &$18.16\pm0.09$ & --- & --- \\
Zeiss-1000 & 2019/11/24& --- & --- & --- &$18.60\pm0.18$ &$18.28\pm0.06$ & $18.18\pm0.11$ & --- \\
BTAS & 2020/01/18& --- & ---& --- & $18.61\pm0.21$ &$18.38\pm0.19$& --- & --- \\
Zeiss-1000 & 2020/02/17& --- & --- & --- &--- &$18.46\pm0.14$ & $18.20\pm0.05$ & $18.17\pm0.13$\\
2.5m CMO & 2020/03/07 & --- & --- & $17.72\pm0.13$&$18.51\pm0.09$ & $18.32\pm0.07$& $18.14\pm0.06$&$18.02\pm0.11$\\
Zeiss-1000 & 2020/11/14 & --- & --- & ---&$18.60\pm0.16$&$18.28\pm0.10$ & $18.19\pm0.07$ & $18.03\pm0.12$ \\
2.5m CMO & 2020/12/03 & --- & --- & ---&$18.54\pm0.08$ & $18.27\pm0.12$& $18.17\pm0.08$&---\\
2.5m CMO & 2021/03/05 & --- & --- & ---&$18.48\pm0.13$ & $18.23\pm0.05$& $18.13\pm0.07$&---\\
 \hline

\end{tabular}
\label{Tab3}
\end{minipage}
\end{table*}

\begin{table*}
\caption{Results of the photometry in H$\alpha$ and near-IR.} 
\begin{tabular}{ccccc} \hline\hline 
\centering
 & \multicolumn{1}{c}{HST/ACS/WFC} & \multicolumn{1}{c}{HST/ACS/HRC}  & \multicolumn{2}{c}{HST/WFC3/IR}  \\
 & \multicolumn{1}{c}{(2005/11/11)} & \multicolumn{1}{c}{(2006/01/26)} & \multicolumn{2}{c}{(2019/04/04)} \vspace{1ex} \\
Object & F658N, & F658N, &F110W, & F160W, \\ 
& mag & mag & mag & mag \\ \hline
J122810.94+440540.6 &-- & $17.44\pm0.05^m$ & $18.46 \pm 0.03$ & $18.06 \pm 0.04$  \\
J122811.70+440550.9 &-- & $17.41\pm0.05^m$ & $19.35 \pm 0.03$ & $18.87 \pm 0.04$ \\ 
J122817.83+440630.8 &$18.33\pm0.05^m$ &-- & -- & -- \\ 
J122809.72+440514.8 &$17.44\pm0.05^m$ &-- & $17.83 \pm 0.02$ & $17.57 \pm 0.02$ \\ \hline
\end{tabular} 
\label{Tab4}
\end{table*}

\section{Results}
\label{res}

\subsection{J122810.94+440540.6}

\subsubsection{Spectra}
\label{specJ122810}

\begin{figure*} 
\begin{center}
\includegraphics[angle=270, width=0.8\linewidth]{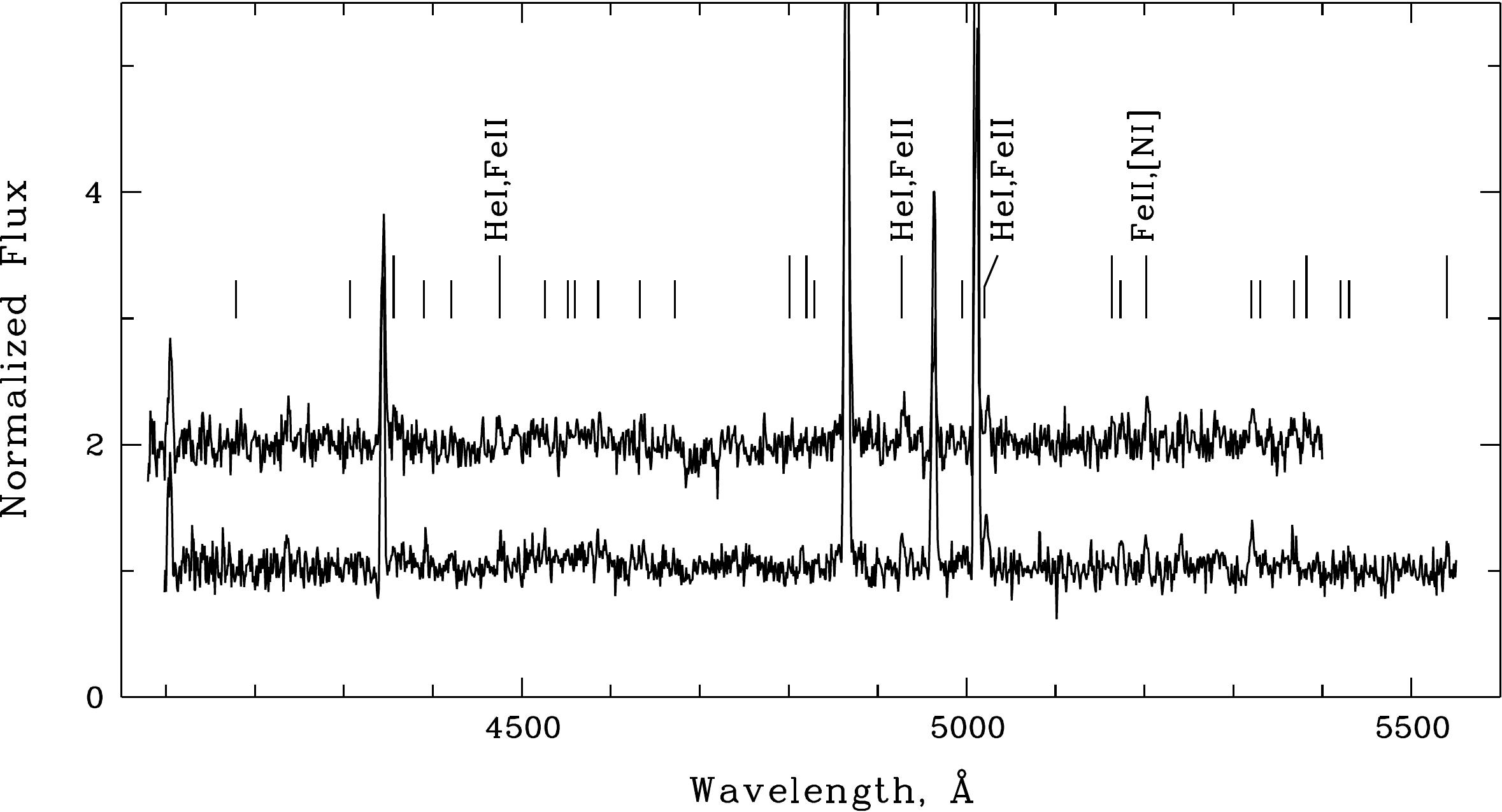}\\
\vspace{5pt}
\includegraphics[angle=270, width=0.8\linewidth]{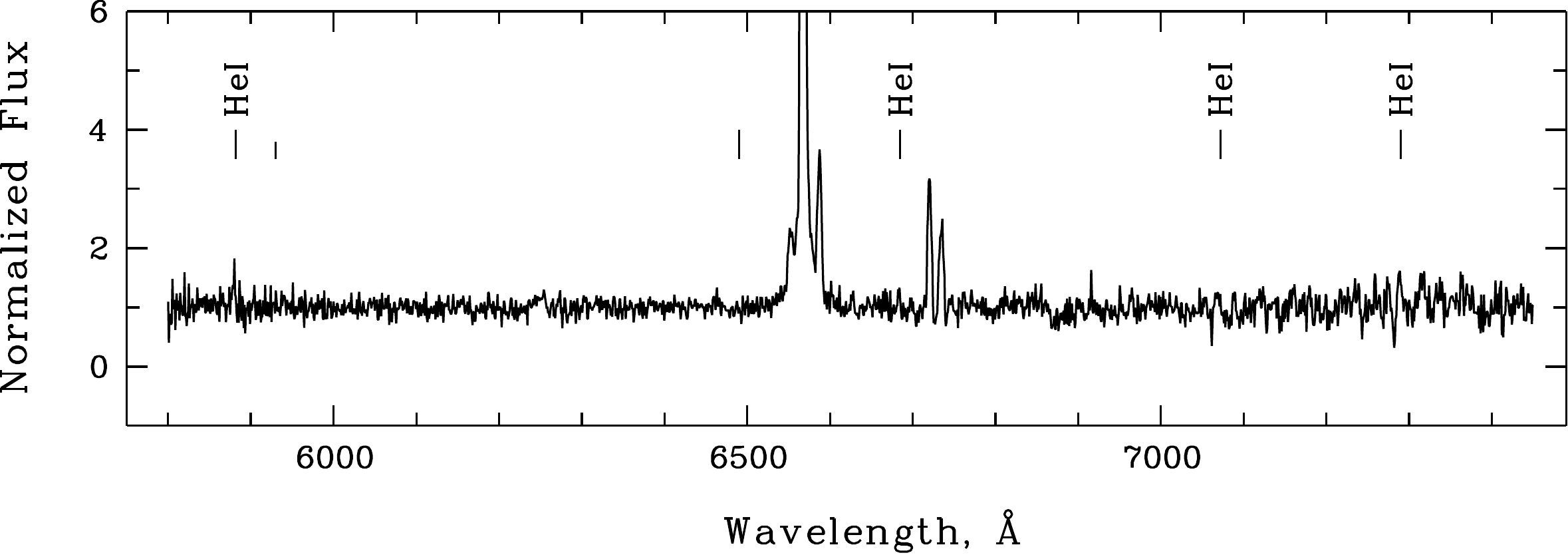}\\
\vspace{1pt}
\caption{Normalized spectra of J122810.94+440540.6, obtained on (\textit{upper panel}) 2015 (bottom) and 2020 (top), and on 2014 (\textit{bottom panel}). The unlabelled short and long ticks represent \ion{Fe}{ii} and [\ion{Fe}{ii}]lines, respectively.} 
\label{Fig2} 
\end{center}
\end{figure*}

The J122810.94+440540.6 spectra obtained with BTA are shown in Fig.\ref{Fig2}. The spectra contain hydrogen Balmer emission lines with obvious broad components. The presence of narrow components in these lines associated with nebula emission makes it difficult to estimate the FWHM of the broad components. The bright narrow emission lines [\ion{O}{iii}]\,$\lambda$4959, $\lambda$5007, [\ion{N}{ii}]\,$\lambda$6548, $\lambda$6583, [\ion{S}{ii}]\,$\lambda$6717, $\lambda$6731 also must belong to the nebula, which is most likely background rather than physically related to the object. The spectra also contain a large number of emission \ion{Fe}{ii}, [\ion{Fe}{ii}] lines and weak \ion{He}{i} lines. There is no significant change in the flux and line shapes between the spectra obtained in 2015 and 2020. We have estimated the interstellar reddening as $A_V=0.2\pm0.2^m$ based on the ratio of hydrogen lines of a nebula measured slightly away from the object assuming case B photoionization \citep{Osterbrock06}.

\subsubsection{Photometry and spectral energy distribution}
 
Photometry of J122810.94+440540.6 did not reveal a significant change in its brightness, the maximal variation is $\Delta U = 0.29\pm0.06^m$, Table~\ref{Tab3}).
 
\begin{figure*}
\begin{center}
\includegraphics[angle=0, width=0.47\linewidth]{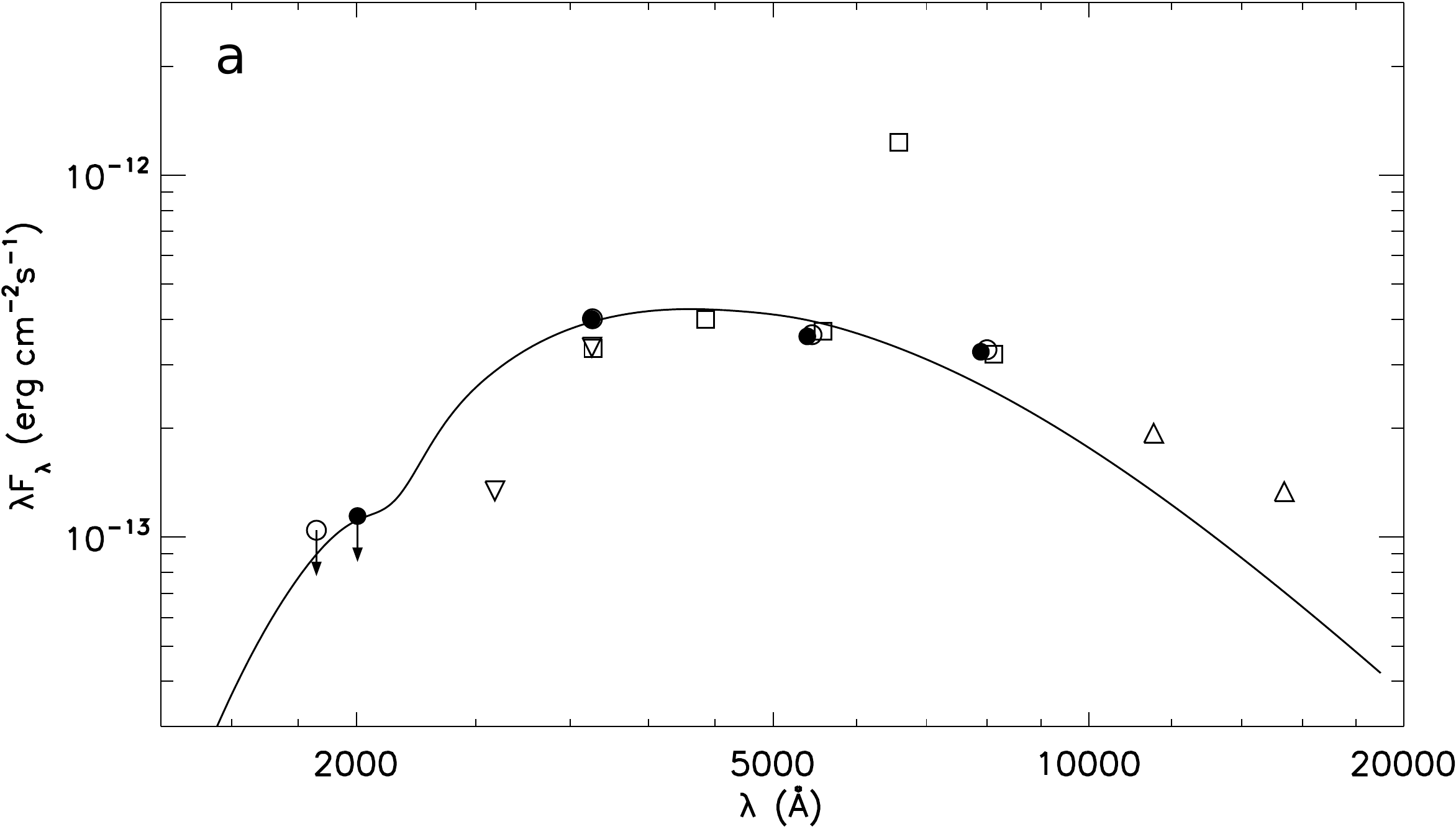}\hspace{5pt}
\includegraphics[angle=0, width=0.47\linewidth]{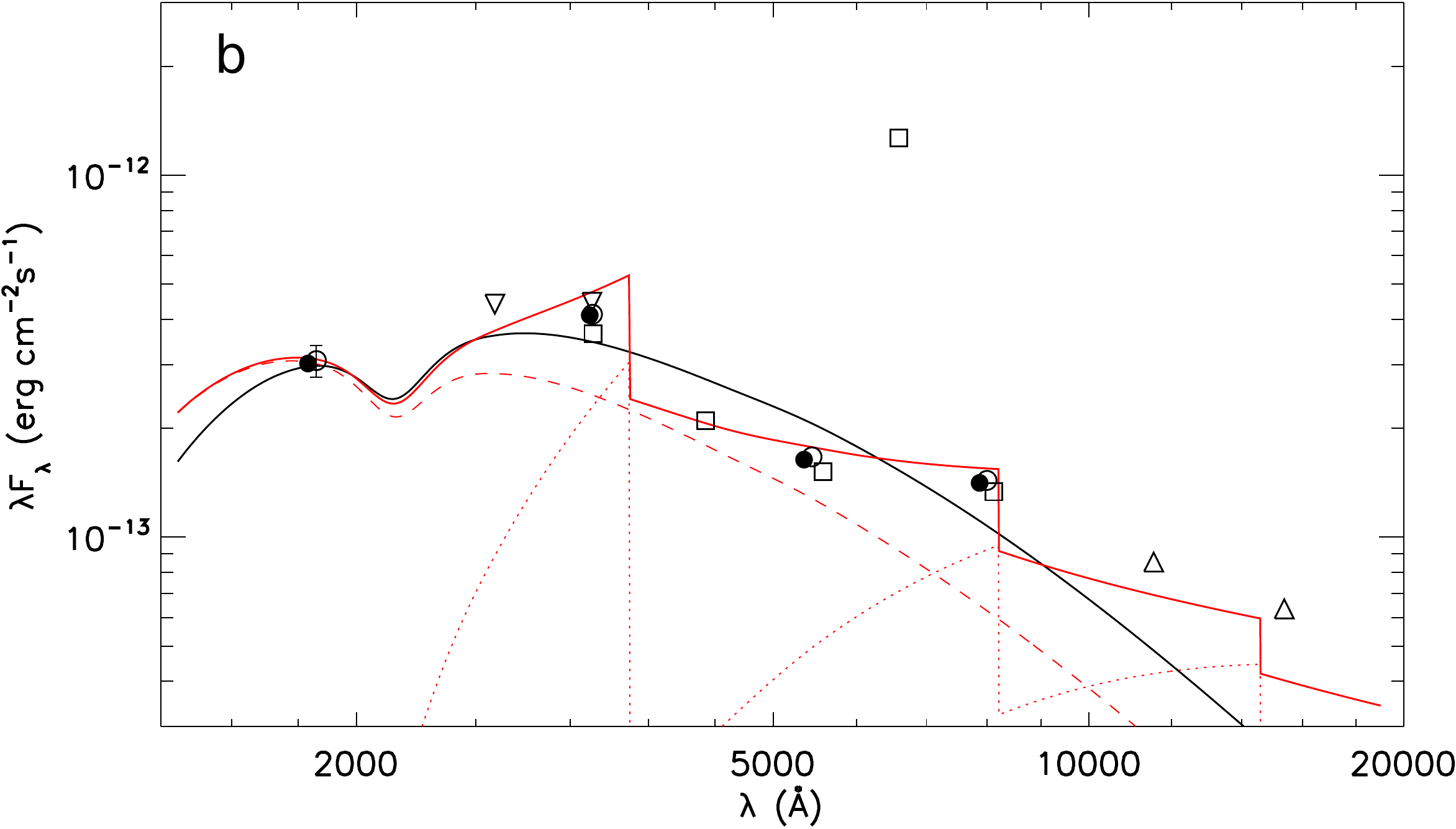}\\
\vspace{5pt}
\includegraphics[angle=0, width=0.47\linewidth]{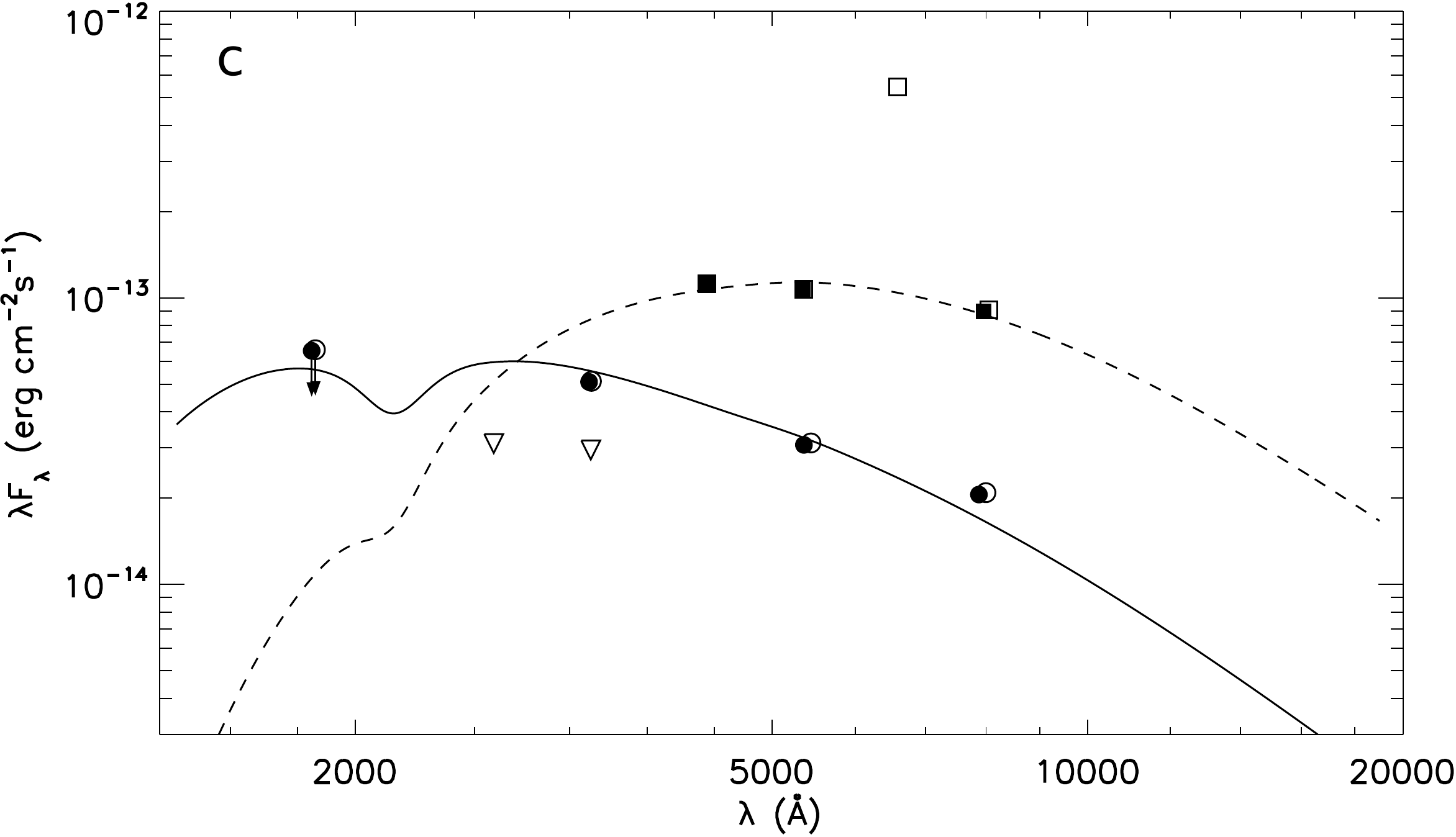}\hspace{5pt}
\includegraphics[angle=0, width=0.47\linewidth]{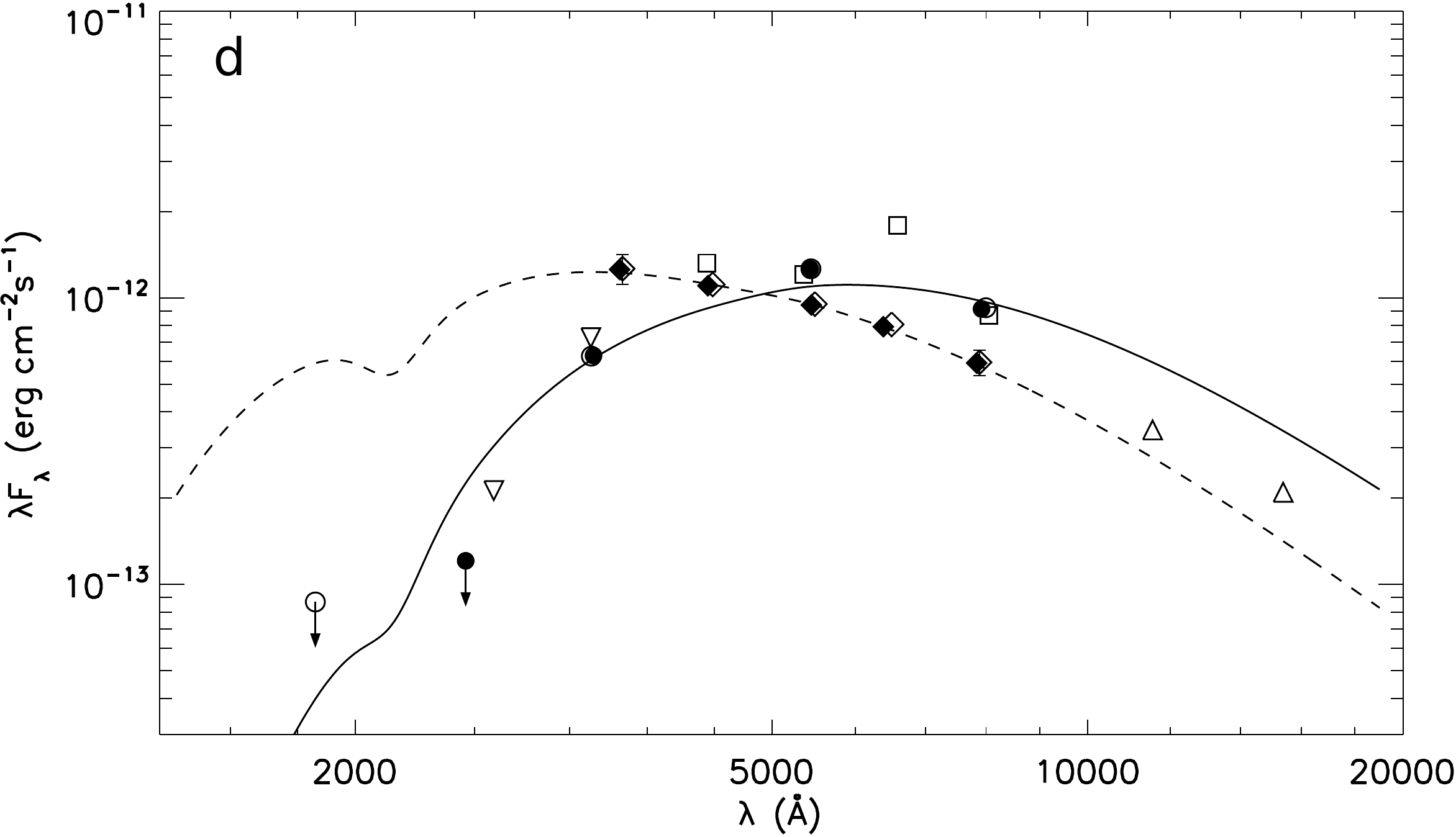}
\caption{SED of the studied objects. The black solid and dashed line represent the best-fitted blackbody model. The unfilled and filled symbols correspond to the data at pivot lambda of \textit{HST} filters and the data corrected for the SED slope during the blackbody fit. The upward triangles represent the IR data from \textit{HST}/WFC3/IR (2019), and the downward triangles represent the UV data from \textit{HST}/WFC3/UVIS (2014) in all figures. (a) J122810.94+440540.6. The squares represent the \textit{HST}/ACS/HRC data (2006). The circles indicate the HST/WFPC2 photometry (1998). (b) J122811.70+440550.9. The squares mark the HST/ACS/HRC data (2006). The circles designate the HST/WFPC2 photometry (1998). The red solid line indicates the composite model including contribution of the blackbody radiation (red dashed line) and f–f and f–b emissions (red dotted line; see details in the text). (c) J122817.83+440630.8. The circles represent the \textit{HST}/WFPC2 photometry (1998). The squares indicate the \textit{HST}/ACS/WFC photometry (2005). (d) J122809.72+440514.8. The circles represent the \textit{HST}/WFPC2 photometry (1997). The squares denote the \textit{HST}/ACS/WFC data (2005). Diamonds mark the data of the 2.5-m telescope of SAI MSU (2020).}
\label{Fig3}
\end{center}
\end{figure*}

The photosphere temperature of J122810.94+440540.6 was estimated from the photometric data in a wide range of wavelengths. For this purpose we fitted the spectral energy distribution (SED) of the object constructed of the magnitudes in original HST filters obtained from the observation of 1998 (Fig.~\ref{Fig3}a). The data of ACS/HRC (2006), IR (2019) and UV (2014) photometry are also shown in Fig.~\ref{Fig3}a but not used during the fitting procedure. The SED demonstrates a downturn in the range of F275W filter, probably associated with strong absorption by metals. The SED was fitted with a blackbody model accounted for interstellar extinction with $R_V=3.07$ \citep{Fitzpatrick99}. The interstellar reddening was restricted to vary within the range $A_V=0.2\pm0.2^m$ derived from the spectroscopy. We have obtained the best-fitting temperature $\rm{T}_{\rm{eff}} = 10000\pm500$K at $A_V \approx 0.4^m$, and the corresponding bolometric magnitude and luminosity $\textrm{M}_\textrm{{Bol}}=-9.60\pm0.23^m$ and $\log(\text{L}_\text{Bol}/\text{L}_{\odot})=5.76\pm0.09$. It is worth noting the presence of notable near-IR excess in F110W and F160W bands, which is probably associated with free-free emission of the wind and/or with the presence of warm circumstellar dust.

\subsection{J122811.70+440550.9}
\subsubsection{Spectra}

\begin{figure*}
\begin{center}
\includegraphics[angle=270, width=0.8\linewidth]{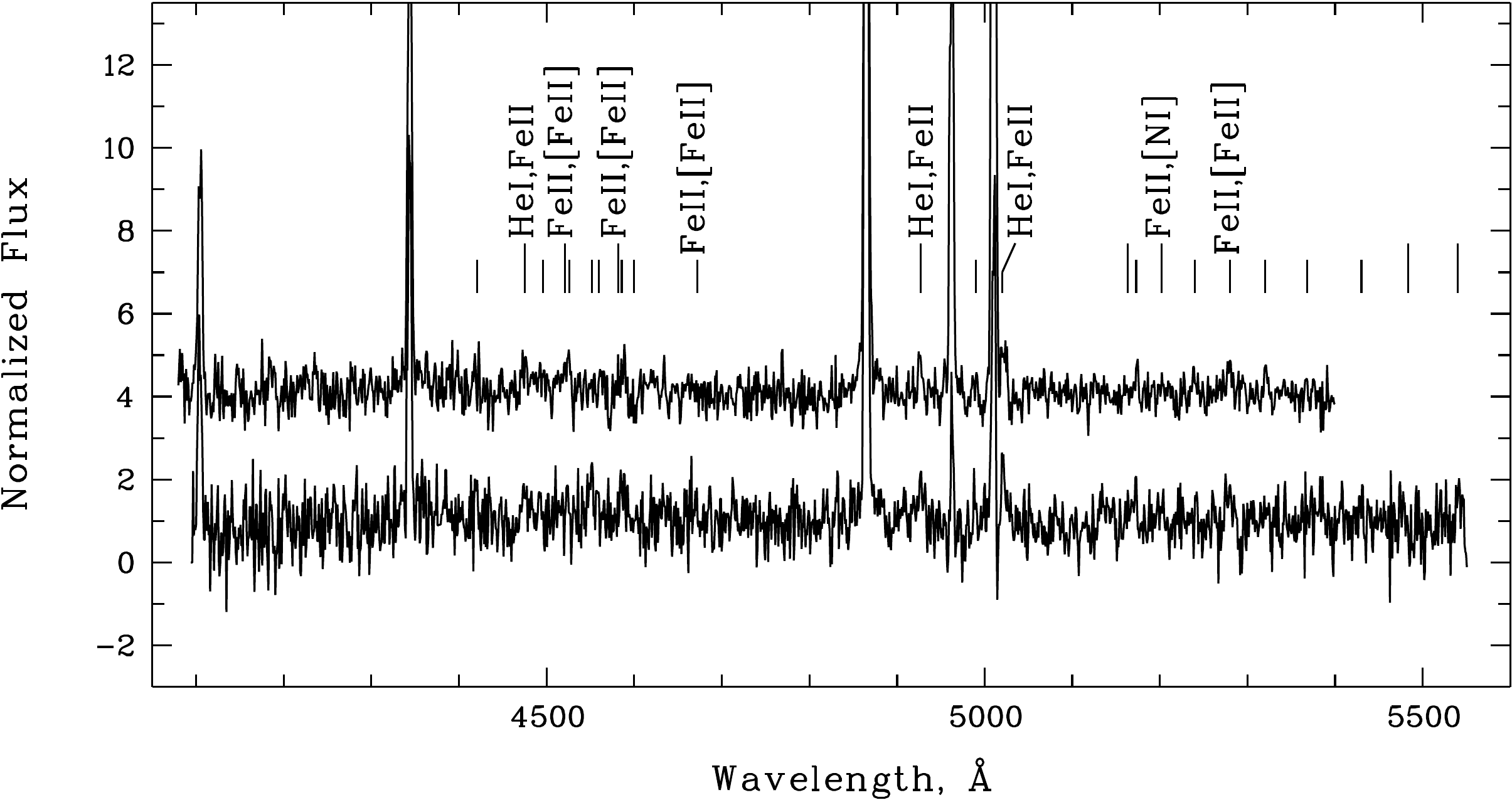}\\
\vspace{5pt}
\includegraphics[angle=270, width=0.8\linewidth]{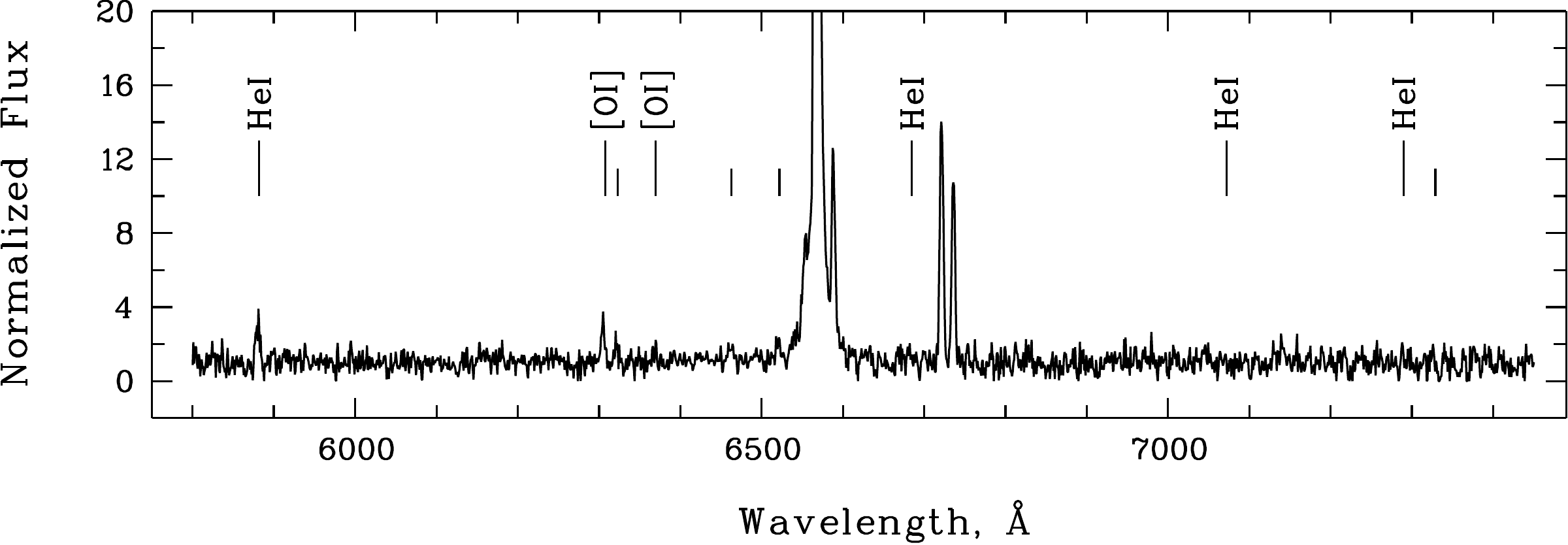}\\
\vspace{1pt}
\caption{Normalized spectra of J122811.70+440550.9, obtained in (\textit{upper panel}) 2015 (bottom) and 2020 (top), and on 2014 (\textit{bottom panel}). The unlabelled short and long ticks represent \ion{Fe}{ii} and [\ion{Fe}{ii}] lines, respectively. Red wing of the nebular [\ion{O}{iii}] line was oversubtracted in the spectrum, taken in 2020.}
\label{Fig4}
\end{center}
\end{figure*}

The spectra of J122811.70+440550.9 (Fig.~\ref{Fig4}) show very broad hydrogen lines H$\alpha$, H$\beta$, H$\gamma$ and H$\delta$, many emission lines of \ion{Fe}{ii}, [\ion{Fe}{ii}], as well as the \ion{He}{i} lines. In addition, the spectrum contains [\ion{O}{i}] $\lambda\lambda$ 6300,6363 lines, however, it is difficult to reveal whether these lines belong to the object or to the surrounding nebula. We do not note significant changes in spectral lines  between the spectra of 2015 and 2020 (Fig.~\ref{Fig4}). Based on the ratio of hydrogen lines emitted by the nebula, we estimated the interstellar reddening as $A_V=0.3\pm0.2^m$.

\subsubsection{Photometry and spectral energy distribution}

The HST photometry did not reveal a noticeable brightness variability of J122811.70+440550.9 (Table \ref{Tab3}): the brightness of of the star is constant within $\approx0.2^m$. 

As in the case of the previous object, we fitted the HST data of 1998 with a black body model with $A_V$ restricted within uncertainties $0.3\pm0.2^m$ obtained from the nebular hydrogen lines. As a result, we have got the best-fitting temperature $\textrm{T}_{\textrm{eff}}=17000\pm7000K$ for $A_V\approx0.5$, and the bolometric magnitude and luminosity estimates $\textrm{M}_\textrm{{Bol}}=-10.00\pm2.00^m$ and $\log(\text{L}_\text{Bol}/\text{L}_{\odot})=5.92\pm0.80$.

The black body model alone gave poor agreement with the observed fluxes (black solid line in Fig.~\ref{Fig3}b) showing a strong discrepancy in ranges of the Balmer and Paschen series limits. This situations is similar to that of the B[e]-supergiant J004415.00 in the galaxy M31 studied by \cite{Sarkisyan2020}. Examining the SED of that object, the authors found a powerful contribution of free–free (f–f) and free–bound (f–b) radiation. This is consistent with the presence of an ionized circumstellar envelope typical for B[e]-supergiants \citep{Zickgraf1985, Zickgraf1986}. Following the methods described by \cite{Sarkisyan2020}, we approximate the observed SED by the black body model with extra spectral components accounting for both f-f and f-b radiation\footnote{When fitting with this more complicated model, additionally to the simultaneous data points of 1998, we utilized the WFC3/UVIS/F275W and ACS/HRC/F435W observations. The choice of these data is explained by the desire to obtain a more detailed SED in order to reach higher accuracy of the model parameters. Taking into account low photometric variability of the object, this choice have to not distort the observed SED.} using \textsc{chianti} package \citep{Dere1997, Landi2013}. We consider the case of isothermal pure hydrogen plasma at a temperature of $\textrm{T}_e = 10000$\,K \citep{Lamers1998}. The reddening was restricted within the uncertainties as above. As a result, we received the best-fitting temperature $\textrm{T}_{\textrm{eff}}=20800\pm4500K$, emission measure of $EM = 1.47\times10^{39}\pm4.37\times10^{38}$ cm$^{-5}$, $A_V\approx0.5$, and the bolometric magnitude and luminosity estimates $\textrm{M}_\textrm{{Bol}}=-9.90\pm1.09^m$ and $\log(\text{L}_\text{Bol}/\text{L}_{\odot})=5.88\pm0.44$\footnote{including only the blackbody component.}

The result of the SED fitting with the composite model is shown in Fig.~\ref{Fig3}b in red lines: the dashed line shows the black body radiation, the dotted line designates f–f and f–b emissions, and the solid red line indicates the total model spectrum. The black solid line represent results of the black body model alone. The IR data (WFC3/IR/F110W,F160W, 2019) and optical data (ACS/HRC/F330W,F550M,F658N,F814W, 2006; WFC3/UVIS/F336W, 2014) are also plotted, but not used in the fit.

\subsection{J122817.83+440630.8}
\subsubsection{Spectra}

The spectrum of the region with J122817.83+440630.8 was obtained with SCORPIO-2 of BTA on 2021 February 11, when the magnitude of the object was $\approx22^m$. During this observations the seeing was about 2.3\arcsec. Thus, the object was too faint for this conditions, and we took only a spectrum of a nearby H\,II complex in order to measure the interstellar extinction. Using the ratio of hydrogen lines, we obtained $A_V=0.4\pm0.2^m$.

\subsubsection{Photometry and spectral energy distribution}

\begin{figure*}
 \begin{center}
 \includegraphics[angle=270, scale=0.45]{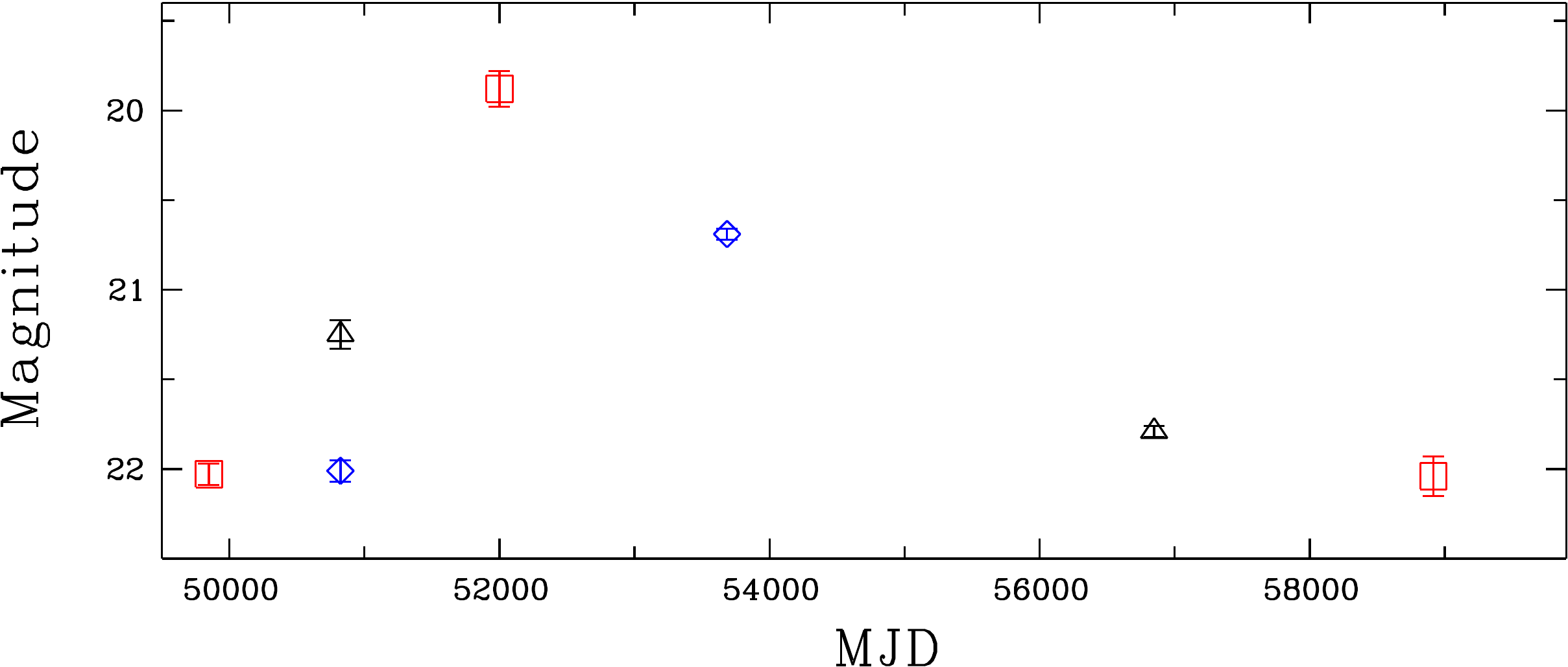}
 \caption{Light curves of J122817.83+440630.8 in \textit{V} (diamonds), \textit{U} (triangles) and \textit{R} (squares) bands.}
 \label{Fig5}
 \end{center}    
\end{figure*}

The most dramatic changes in brightness of this star occurred in the R band (Table~\ref{Tab3}). From the table it is seen that the source brightness increased by $\Delta R=2.15\pm0.13^m$ from the 1995 (HST) to 2001 (Bok) observations, and a return to its previous state to 2020 (2.5-m CMO). The light curves in different filters are shown in Figure \ref{Fig5}.

The shape of the object SEDs also dramatically changed (compare the data of 1998 and 2005 in Fig.~\ref{Fig3}c). To measure the photosphere temperatures, we carried out a joint approximation of these SEDs with a black body model taking into account that the interstellar extinction and the bolometric luminosity of the two data sets have to be the same if the observed variability is of the S Dor-type. As before, $A_V$ was limited to the range measured from the spectroscopy. As a result, we obtained the temperatures $\textrm{T}_{\textrm{eff}}=19000\pm1200$K (1998) and $\textrm{T}_{\textrm{eff}}=9000\pm600$K (2005) for the interstellar extinction $A_V\approx0.6^m $, which gives estimates of the bolometric magnitude $\textrm{M}_\textrm{{Bol}}=-8.30\pm0.38^m$ and the corresponding bolometric luminosity $\log(\text{L}_\text{Bol}/\text{L}_{\odot})=5.24\pm0.15$.

\subsection{J122809.72+440514.8}
\label{Spec_J122809}
\subsubsection{Spectra}

\begin{figure*}
\begin{center}
\includegraphics[angle=270, width=0.8\linewidth]{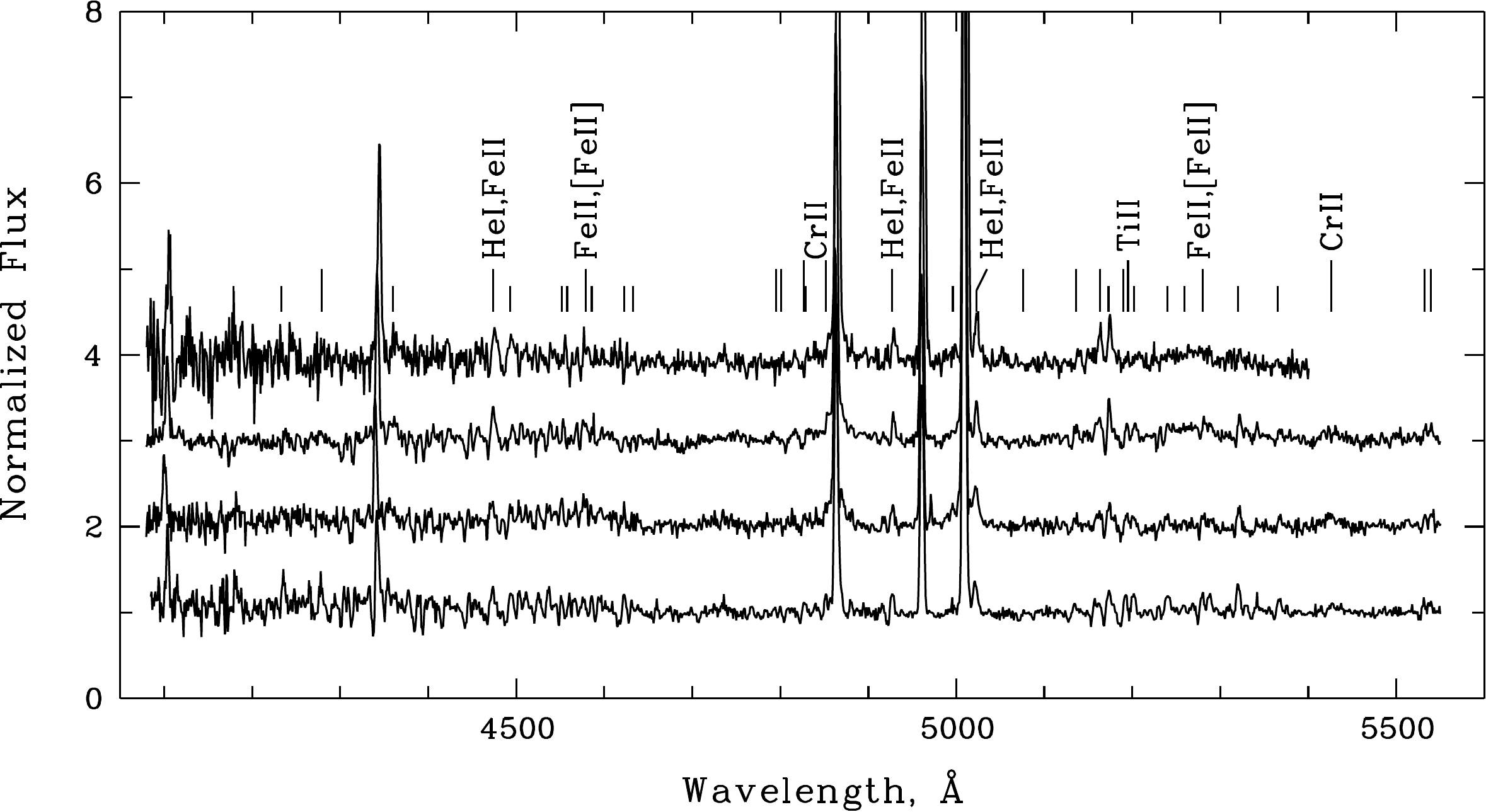}\\
\vspace{5pt}
\includegraphics[angle=270, width=0.8\linewidth]{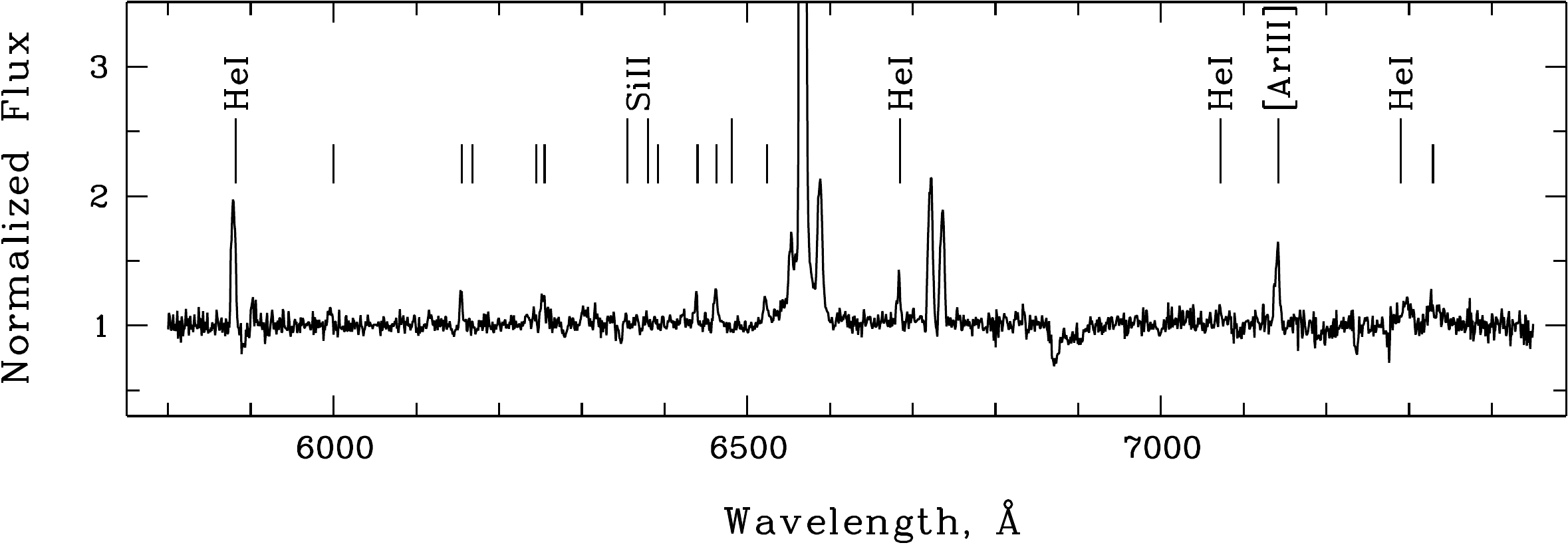}\\
\vspace{1pt}
\caption{Normalized spectra of J122809.72+440514.8. The spectra were obtained in 2015, 2017, 2018 and 2020 (from bottom to top; \textit{upper panel}), in 2014 (\textit{bottom panel}).}
\label{Fig6}
\end{center}
\end{figure*}

The spectra of J122809.72+440514.8 are shown in Fig.~\ref{Fig6}.
The spectrum obtained in 2015 contains emission lines \ion{He}{i} $\lambda\lambda$4472,4922 and \ion{Fe}{ii} $\lambda\lambda$4400-4700, $\lambda\lambda$5100-5400 with P Cyg profiles which became less noticeable in 2017 and 2018, and then completely disappeared by 2020.
The wings of the hydrogen lines H$\beta$ and H$\gamma$ broadened noticeably from 2015 to 2020; the [\ion{Fe}{ii}] $\lambda$5157 and \ion{Fe}{ii} $\lambda$5169 lines become brighter. At the same time, there is a notable weakening of other emission lines of \ion{Fe}{ii} and [\ion{Fe}{ii}] $\lambda\lambda$4500-4700 and $\lambda\lambda$5200-5400. In addtition, the spectra show a few \ion{Cr}{ii} and \ion{Ti}{ii} lines. The emission lines \ion{He}{i} $\lambda$ 5876, $\lambda$6678 are also seen but their FWHM correspond to the spectral resolution, so these lines are probably emitted by the surrounding nebula.
The study of the hydrogen lines of the surrounding nebula allowed us to estimate the interstellar extinction $A_V=0.8\pm0.2^m$.

\subsubsection{Photometry and spectral energy distribution}\label{sed_J122809.72+440514.8}

\begin{figure*}
    \begin{center}
    \includegraphics[angle=270, scale=0.45]{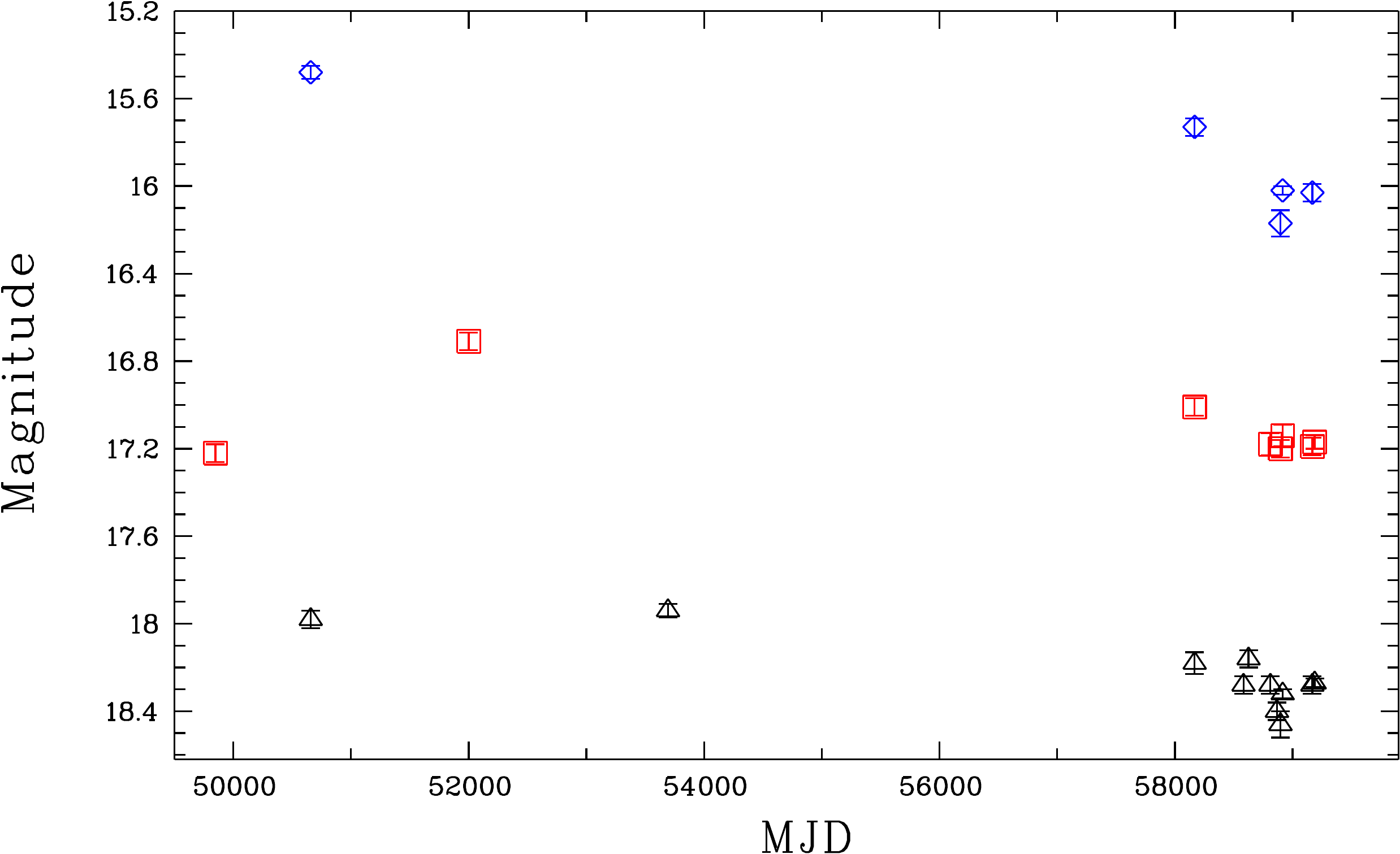}
    \caption{Light curves of J122809.72+440514.8 in \textit{V} (triangles), \textit{R} (squares) and \textit{I} (diamonds) bands. To better visualisation light curves in \textit{R} and \textit{I} band shifted on -1 and -2 mag respectively.}
    \label{Fig7}
    \end{center}
\end{figure*}

The brightness variations of J122809.72+440514.8 are $\Delta U=0.28\pm0.24^m$, $\Delta V=0.48\pm0.14^m$, $\Delta I = 0.69\pm0.13^m$ from 1997 (HST) to 2020 (Zeiss-1000, 2.5-m CMO). The light curve is shown in Fig.~\ref{Fig7}.

Monitoring of the star with ground-based telescopes carried out over the past three years, has revealed the strongest changes in the star colour and brightness. The earlier data (HST) show weaker variability, and, at the same time, demonstrate (up to 2014 inclusively) a clear downturn in the range 3000-4000\AA\ (Fig.~\ref{Fig8}), which we identify as a quite deep Balmer jump. The observations of 2020 in a wide range of wavelengths (from the U to I band) show a significantly hotter SED without clear Balmer jump.

In the case of J122809.72+440514.8, the joint approximation of two extreme states of the star (we used the data of 1997 from HST and of 2020 from 2.5-m CMO) with a black body model and tied $A_V$ and $\text{L}_\text{Bol}$ values of the two data set, as we carried out for the previous object, did not give satisfactory results. The reason for this could be variations either in the absorption value or in the bolometric luminosity. The change of $A_V$ was observed in $\eta$~Car, associated with condensation of dust in the shell ejected during the Great Eruption \citep{DavidsonHumphreys1997}. However, we did not find in the literature any examples of absorption changes in the environment around LBVs undergoing the S Dor cycle.
The bolometric luminosity changes, nevertheless, have been observed in several confirmed LBVs during their S Dor-type variability cycle. The example is AG Car, whose bolometric luminosity decreased by a factor of 1.5 when the source was at the brightness maximum in the V band \citep{Groh09}. A similar behaviour of S Dor was noted by \cite{Lamers95}. They also has shown that the bolometric luminosity of LBVs can vary within 0.2 dex. This may occur due to the loss of energy for the expansion of the envelope during transitions of the star brightness from minimum to maximum.

In accordance to the above, we repeated the fit with the bolometric luminosity being free but the extinctions for the two SEDs were still tied and restricted within uncertainties. The best-fitting models gave $\textrm{T}_{\textrm{eff}}=7600\pm300$~K,  $\log(\text{L}_\text{Bol}/\text{L}_{\odot})=6.20\pm0.11$ ($\textrm{M}_{\textrm{Bol}}=-10.70\pm0.27^m$) for the `cold' state (1997), and $\textrm{T}_{\textrm{eff}}=13500\pm4300$,  $\log(\text{L}_\text{Bol}/\text{L}_{\odot})=6.44\pm0.64$ ($\textrm{M}_{\textrm{Bol}}=-11.30\pm1.60^m$) for the `hot' state (2020), with $A_V=0.6^m$. Both SEDs together with corresponding models are shown in Fig.~\ref{Fig3}d.

The difference between the two obtained values of $\text{L}_\text{Bol}$ lies within the 0.2~dex limit discussed above. However, it is still possible that this variation of $\text{L}_\text{Bol}$ is an artefact associated with underestimation of the photosphere temperature in the cold state due to the presence of the Balmer jump. To test this assumption, we decided to use the CMFGEN non-LTE models \citep{Hillier98}, which are capable to give more accurate estimates of the fundamental stellar parameters under conditions of a strong gas outflow in the form of a wind.

\subsubsection{CMFGEN model}

The photometric data taken in 2014 in the F275W and F336W filters of the HST/WFC3/UVIS are in a good agreement with the UV-range data of 1997, which indicates the absence of significant difference in the states of the star between these observation dates. The spectral features in the first spectrum of the star, which we obtained in 2015, are consistent with relatively low temperatures of the outflowing gas. Therefore, we assumed that this spectrum also corresponds to the cold state of the star, observed at least in 1997 and 2014, and we decided to use it for modelling with the \textsc{cmfgen} code. Beside the spectrum, we involved the photometric data of these years to obtain estimates of the star luminosity.

As a first approximation of the J122809.72+440514.8 spectrum, we utilised the models from our previously calculated CMFGEN grids of extended atmosphere models (Kostenkov et al. 2021, in prep.). A detailed description of the modelling method and algorithm is given in the works \citet{Kostenkov20a, Kostenkov20b}. The velocity distribution in the stellar wind was assumed to obey a simple velocity law with $\beta=1$ \citep{Lamers1996}. This value of $\beta$ provides a very rapid increase of the gas velocity near the star surface, which leads to the compact photosphere that follows from the presence of the absorption Balmer jump observed in the star SED. The absence of forbidden lines formed in distant wind regions (for example, [\ion{N}{ii}] $\lambda5755$) does not allow us to accurately estimate the terminal wind speed. Therefore, we have accepted its value equal to 300 km s$^{-1}$, which made it possible to reproduce the P Cyg profiles of iron lines quite accurately at the given velocity law. Since we can not observe the electron-scattering wings of the hydrogen lines due to the low signal-to-noise ratio, we assume homogeneity of the wind structure (filling factor $f=1$). The metals abundance is assumed to be equal to the metallicity of NGC\,4449 $Z=0.5Z_{\odot}$ \citep{Annibali2017}, with the exception of the nitrogen and carbon abundance that was changed to be consistent with the values observed in detailed studied LBVs (Table~\ref{Tab5}). The hydrogen abundance was determined iteratively based on the ratio [\ion{O}{iii}]$\lambda$5007/H$\beta$: we subtracted from the H$\beta$ its stellar component obtained from the model of current iteration, and sought to make the ratio of [\ion{O}{iii}]$\lambda$5007 and the `H$\beta$ remnant` the same as observed in the pure nebula spectrum extracted from nearby region.

\begin{table}
    \centering
    \caption{The main model parameters and chemical abundances ($X$) for J122809.72+440514.8.}
\begin{tabular}{ | p{130pt} | r | }
\hline
$L_{*}$, $L_{\odot}$ & 2.58 $\times$ $10^6$   \\ 
$\dot{M}$, $M_{\odot}$ yr$^{-1}$ &  5.2 $\times$ $10^{-3}$ \\
$R_{2/3}$ $^{*}$, $R_{\odot}$ & 620  \\
$R_{*}$ $^{*}$, $R_{\odot}$ & 410  \\
$T_{\text{eff}}$ $^{*}$, K & 9300  \\
$T_{*}$ $^{*}$, K & 11360 \\
$\beta$ & 1.0\\
$V_{\infty}$, km s$^{-1}$  & 300\\
$f$ & 1 \\
H, \% & 20\\
$X_{\text{C}}/X_{\odot}$ & 0.2\\
$X_{\text{N}}/X_{\odot}$ & 5.4\\
$X_{\text{Si}}/X_{\odot}$ & 0.5\\
$X_{\text{Fe}}/X_{\odot}$ & 0.5\\
\hline
    \end{tabular}
    \label{Tab5}
    
    \textit{Notes.} $^{*}$ $R_{2/3}$ and $T_{\text{eff}}$ are the radius and temperature at $\tau = 2/3$, $T_{*}$ is the temperature at hydrostatic radius $R_{*}$ ($\tau \gtrsim 20$)
\end{table}

A relatively good agreement between the observed spectra and the model was reached at $\dot{M} = 5.2\times10^{-3}\,M_{\odot}\,yr^{-1}$ and the temperature $T_{*} = 11360$\,K at the hydrostatic radius of the star $R_{*} \approx 410\,R_{\odot}$ (Fig.~\ref{Fig9}). The main model parameters are presented in Table~\ref{Tab5}. The extremely powerful outflow appearing in this model, increases opacities at large distances from $R_{*}$ and significantly decreases the temperature at the photosphere radius, $T_{\text{eff}} = 9300$\,K. However, this estimate of the effective temperature still significantly exceeds the value obtained from the black body approximation of the SED ($\textrm{T}_{\textrm{eff}}=7600\pm300$K), which indicates underestimation of the temperature by the simple model.

The accuracy of the the mass loss rate estimates is limited by the uncertainty of the nebula contribution to the  hydrogen lines observed in the object spectrum. This uncertainty could be reduced by further observations with a higher spectral resolution. The accuracy of the temperature estimates is determined by several factors. First of all, it is the absence of \ion{Si}{ii} lines in the spectrum, which are sufficiently sensitive to changes in the ionization state of the wind matter. Therefore, the upper limit of the photosphere temperature in the model was limited to $\sim11000\,$K. The lower limit for the effective temperature ($\sim 9000\,$K) was estimated based on the strength of the absorption components of the H$\beta$ and \ion{Fe}{ii} lines. A decrease in temperature leads to an increase in the amount of neutral hydrogen, and hence to an increase in the absorption component of the H$\beta$ line. Another consequence of a decrease in temperature below $\approx9000\,$K is a strong weakening of the \ion{Fe}{ii} lines. In principle, an increase in the mass loss rate can compensate weakening of ionized iron lines; however, the resulting increase in the wind density also strengthens the absorption components of these lines. Taking into account that strong absorption lines are not observed in the object spectrum, we assume that the lower estimate of the photosphere temperature is $9000\,$K.

\begin{figure*}
 \centering
 \includegraphics[angle=0, scale=0.23]{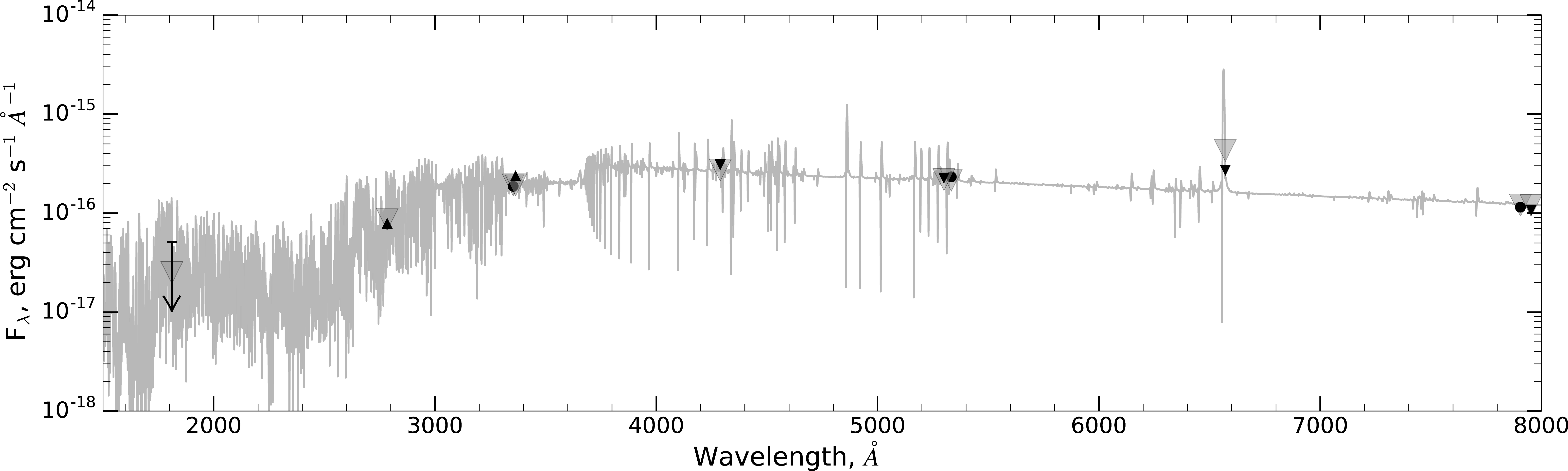}
 \caption{The best-fetting model to J122809.72+440514.8 spectrum, obtained on 2015 (grey solid line). Fluxes in the F170W, F336W, F555W and F814W of the best-fitting model are marked by grey triangles. The black circles designate photometric data from \textit{HST}/WFPC2 (1997), black downward triangles - data from \textit{HST}/ACS/WFC (2005), black upward triangle - WFC3/UVIS data (2014).}
    \label{Fig8}
\end{figure*}

\begin{figure*}
 \begin{center}
 \includegraphics[angle=0, scale=0.245]{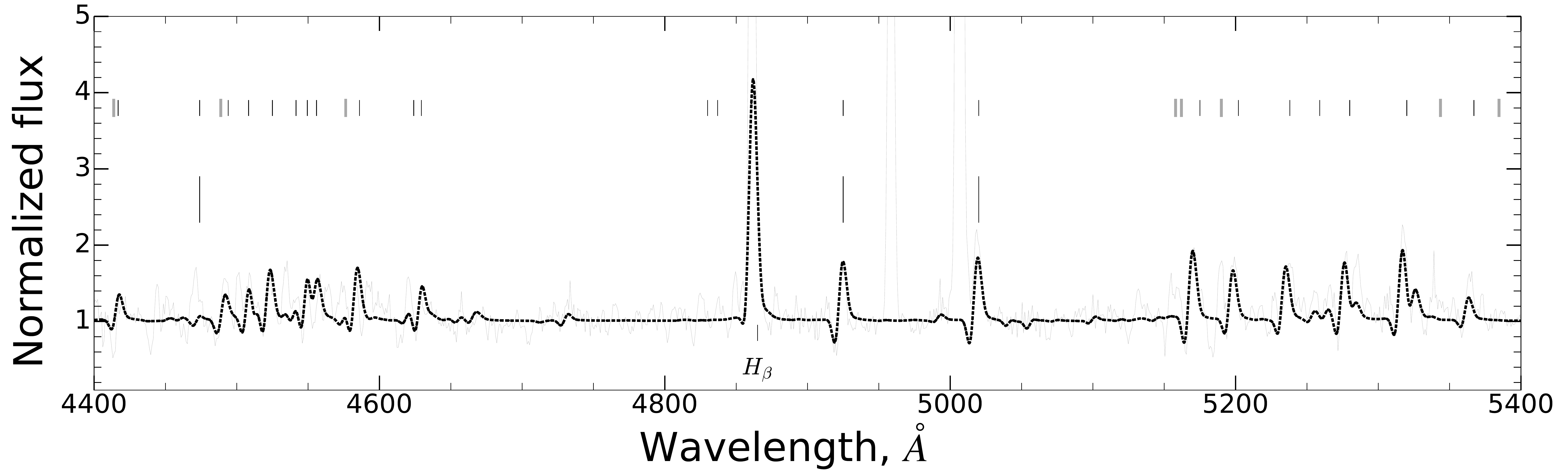}
 \caption{Normalized spectrum of J122809.72+440514.8, obtained in 2015 (grey solid line) and the best-fitting \textsc{cmfgen} model (black solid line). Long black ticks designate \ion{He}{i} lines, \ion{Fe}{ii} and [\ion{Fe}{ii}] are marked with short black and thick grey ticks, respectively}
 \label{Fig9}
 \end{center}   
\end{figure*}

The bolometric luminosity was determined by fitting of the obtained model spectrum accounted for the interstellar extinction to the observed SED (1997).
In this case, the extinction $A_V$ was a free parameter. The best agreement between the observed energy distribution and the model was obtained at $A_V=1.05\pm0.07^m$, which corresponds to the value obtained from observations, but is much higher than the estimate based on the black body fit ($0.6^m$). The luminosity estimate is $\log(\text{L}_\text{Bol}/\text{L}_{\odot})=6.41\pm0.03$. The model together with the observed SEDs are given in Fig.~\ref{Fig8}, where the observational data are shown in black and the model fluxes in grey. The figure shows the observed fluxes from the ACS/WFC data (2005) and the WFC3/UVIS data (2014). Obvious agreement between all the observed and calculated fluxes confirms the initial assumption that the star was in approximately the same state in 1997, 2005 and 2014-2015.

То summarize, we carried out a modeling of the optical spectrum of the cold state of J122809.72+440514.8. However, the poor quality of the available data, its contamination by the nebula lines as well as the absence of IR- and UV-spectra did not allow us to properly constrain such important parameters as $\beta$, the terminal velocity, the hydrogen abundance, etc., which, in turn, made less accurate the estimates of the mass loss rate and photosphere temperature. Nevertheless, our modeling is able to reproduce many observed features both in the spectrum and in the SED.

\section{Discussion}

\subsection{Spectral classification}

As is noted in Sec.~\ref{specJ122810}, the spectrum of J122810.94+440540.6 contains \ion{Fe}{ii}, [\ion{Fe}{ii}] lines, but has no [\ion{O}{i}] $\lambda\lambda$ 6300,6363 and [\ion{Ca}{ii}] $\lambda\lambda$ 7291,7323 lines, which are indicators of circumstellar gas \citep{Aret12} and can be observed in spectra of B[e]-supergiants and warm hypergiants \citep{Humphreys13, Humphreys14}. At the same time, the \ion{Fe}{ii} and [\ion{Fe}{ii}] lines can be observed in stars of various types, including LBVs. The SED of this object shows a noticeable IR excess in the region of 1-2 $\mu$m (Fig.~\ref{Fig3}a). However, the excess is not large (less than 2 times higher than the black body model in the F160W band, approximately corresponding to the H band), therefore, it is not possible to unambiguously establish the source of this excess without mid- and far-IR data: the possible origin of the observed excess can be both a free-free wind radiation and a warm dust circumstellar envelope. Also, J122810.94+440540.6 did not show any noticeable brightness variability (it is less than 0.3$^m$ in all the studied bands). Considering all of the above facts, we cannot clearly classify the star, but we suppose that J122810.94+440540.6 could be either a LBV or B[e]-supergiant.

According to the spectral features, the star J122811.70+440550.9 is similar to J122810.94+440540.6, however, its spectrum contains the [\ion{O}{i}] $\lambda\lambda$ 6300,6363 lines, but there are no obvious [\ion{Ca}{ii}] lines. Lines of neutral oxygen can be excited both in the envelope of the star itself and in the gas of the surrounding nebula. The SED of this object is found to be similar to those of B[e]-supergiant J004415.00 \citep{Sarkisyan2020}. Applying the same model as was proposed for that B[e]-supergiant, which takes into account f-f and f-b radiation, to the SED of J122810.94+440540.6, we obtained a good agreement of the model with the observed data and have demonstrated the presence of an ionized envelope around the star. Thus, J122811.70+440550.9 shows characteristics typical of B[e]-supergiants. However the fragmentariness of the available data do not allow us to draw definitive conclusions.

The star J122817.83+440630.8 demonstrated significant brightness ($\Delta R=2.15\pm0.13^m$) and SED shape variability. The lack of spectral data does not allow to make unambiguous conclusion concerning the object type. Nevertheless, we have to note that such brightness changes are characteristic only for LBVs, since only they are highly variable objects among the entire set of high luminosity stars \citep{Humphreys17}. Various transient X-ray binaries are another type of objects with a large amplitude of optical variability and a hot photosphere. However, we examined images of Chandra and found no X-ray sources at the J122817.83+440630.8 position. So it is unlikely that the object is one of them. Thus, this object can probably be attributed to LBV stars.

The spectrum of J122809.72+440514.8 contains many lines characteristic of LBVs. This star also demonstrated spectral and photometric variability, the maximum of which is in the red range $\Delta I=0.69\pm0.13^m$. The SED of this object shows no clear IR excess, which indicates the absence of a hot gas and dust envelope around the star. Based on the observed features, J122809.72+440514.8 can be classified as an LBV star.

The spectrum of J122809.72+440514.8, obtained in 2015, shows similarity with the spectra of many confirmed LBVs in the cold state (e.g. HR\,Car, \citealt{Szeifert2003}; AG\,Car, \citealt{Stahl2001}; Var~C \citealt{Humphreys2014b}) but it is especially similar to the spectrum of $\eta$ Car of 2001 (Fig.~7 in \citealt{Groh2012}). The authors performed a quantitative analysis of $\eta$ Car spectrum and its modeling using the CMFGEN code. 

The results of modelling of the J122809.72+440514.8 spectrum gave extremely high values of the mass loss rate $\dot{M}=5.2\times10^{-3}\,M_{\odot}\text{/yr}$, which is higher but still comparable to that of $\eta$ Car ($\dot{M}=2.4\times10^{-3}\,M_{\odot}\text{/yr}$, \citealt{Groh2012}). 
Nevertheless, we should note that it is difficult to assess the reliability of the obtained value because there is a degeneracy between the mass loss and the ratio of helium and hydrogen abundances at low temperatures: an increase in the hydrogen content can be compensated for by a corresponding increase in $\dot{M}$. The same difficulty arose when modeling the $\eta$ Car spectrum \citep{Hillier2001}.  
Moreover, the use of other spectral lines can not help to avoid the degeneracy.
For example, a decrease in $\dot{M}$ can lead to a decrease in the model intensities of the \ion{Fe}{ii} lines but, at the same time, the decrease in the intensities of these lines can be compensated for by an increase in the \ion{Fe}{ii} abundance, which is not known precisely due to the uncertainty in the metallicity of the star. Nevertheless, despite all the difficulties, we can assert that both stars have very powerful winds with a large mass loss in them.

$\eta$ Car has a significantly larger photosphere ($\approx 860\,R_{\odot}$) than J122809.72+440514.8 ($\approx 620\,R_{\odot}$) with almost the same temperature and with lower mass loss rate. We suppose that this is due to the following factors: firstly, to the significantly different selected values of the clamping, and secondly, due to the use in \cite{Groh2012} of a more complex velocity law, derived from the available spectral data in a very wide wavelength range (far UV to IR range). In the case of J122809.72+440514.8, there were not enough spectral lines in the optical range, so we applied the simplest model of the $\beta$-law. The less pronounced Balmer jump indicates a more extended photosphere of $\eta$ Car compared to J122809.72+440514.8.

\subsection{Age of the environment stars}

%Fig.11
\begin{figure*} 
\center{
\includegraphics[angle=0, width=0.3\linewidth]{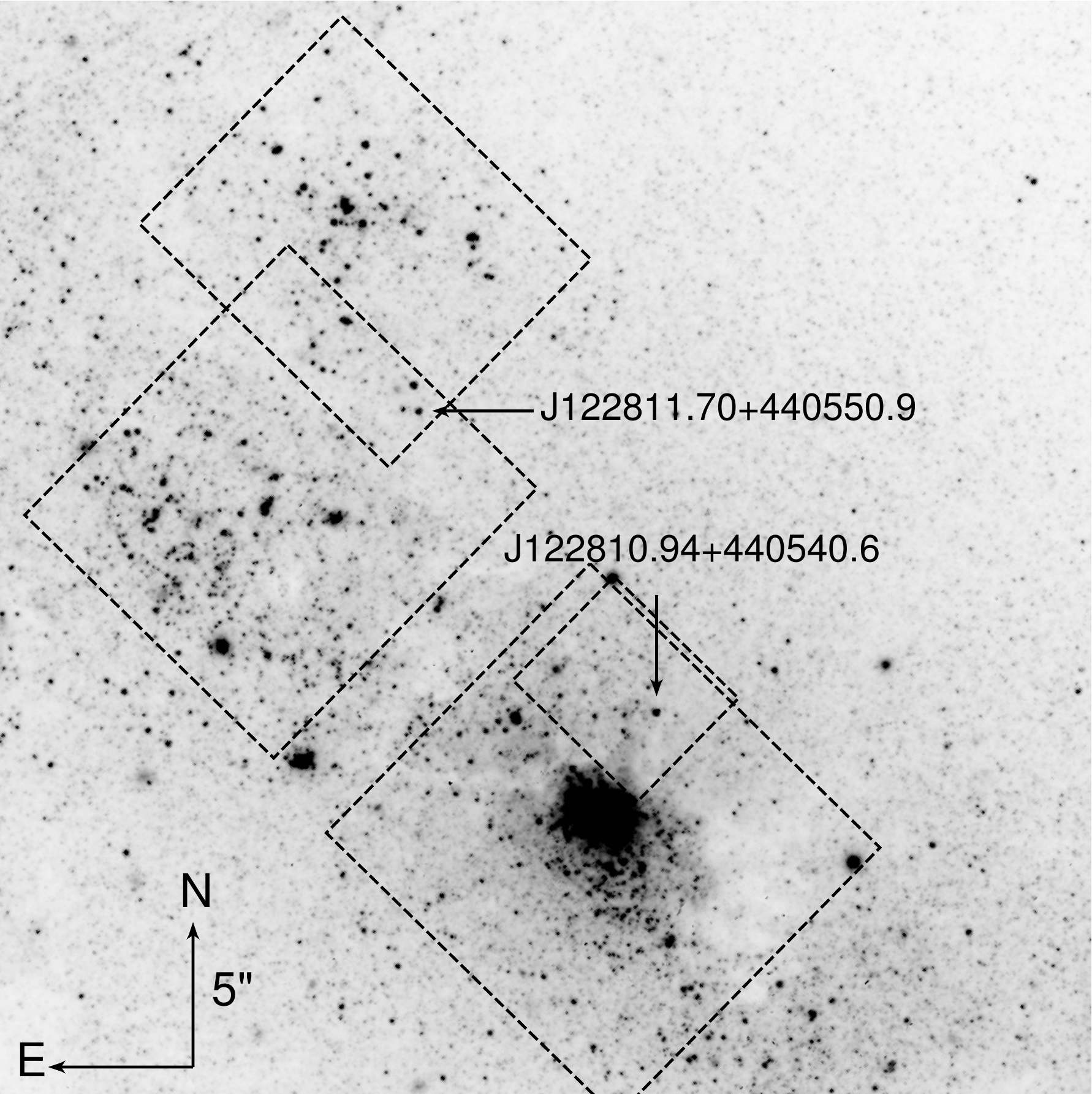}
\includegraphics[angle=0, width=0.3\linewidth]{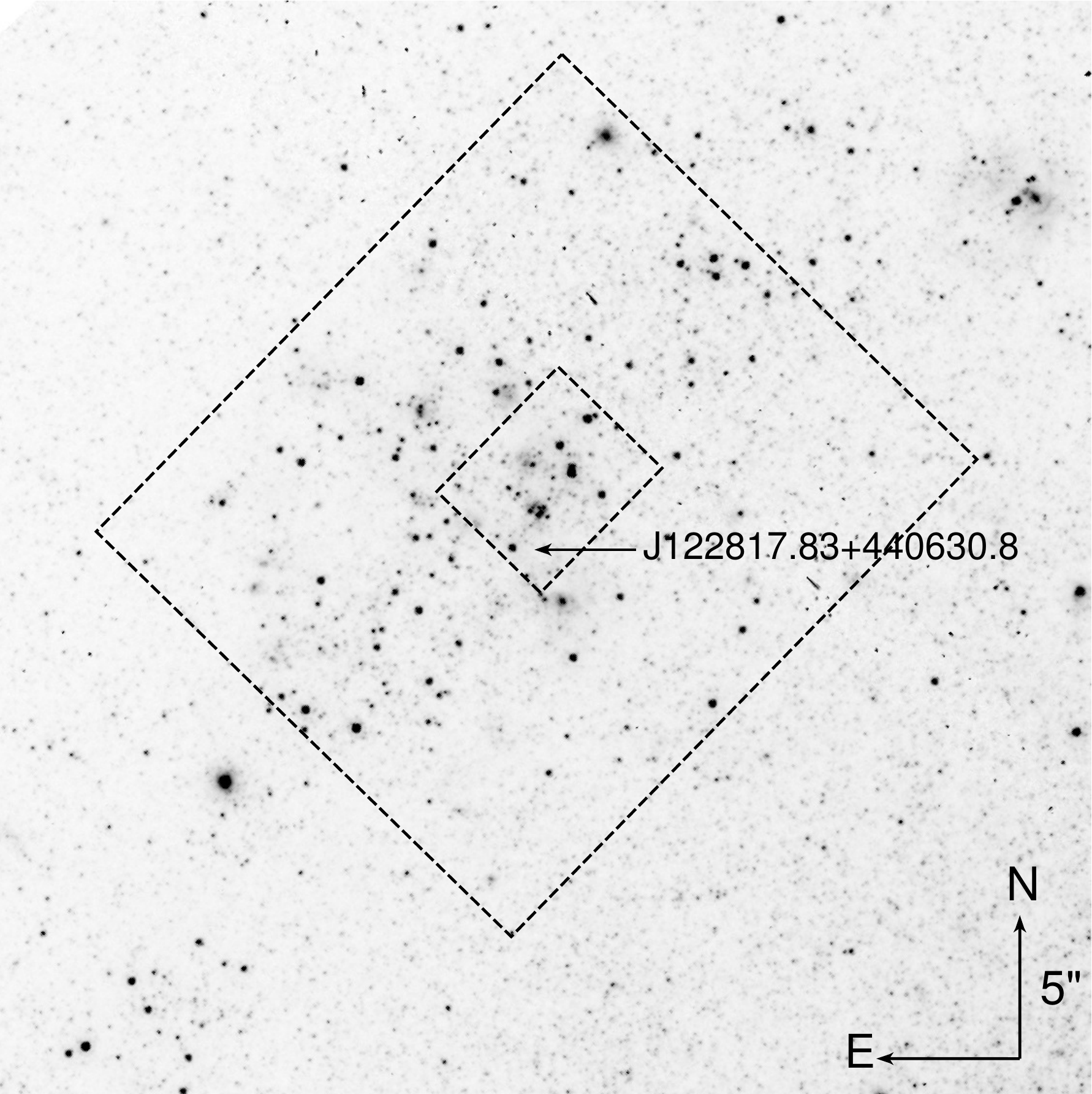}
\includegraphics[angle=0, width=0.3\linewidth]{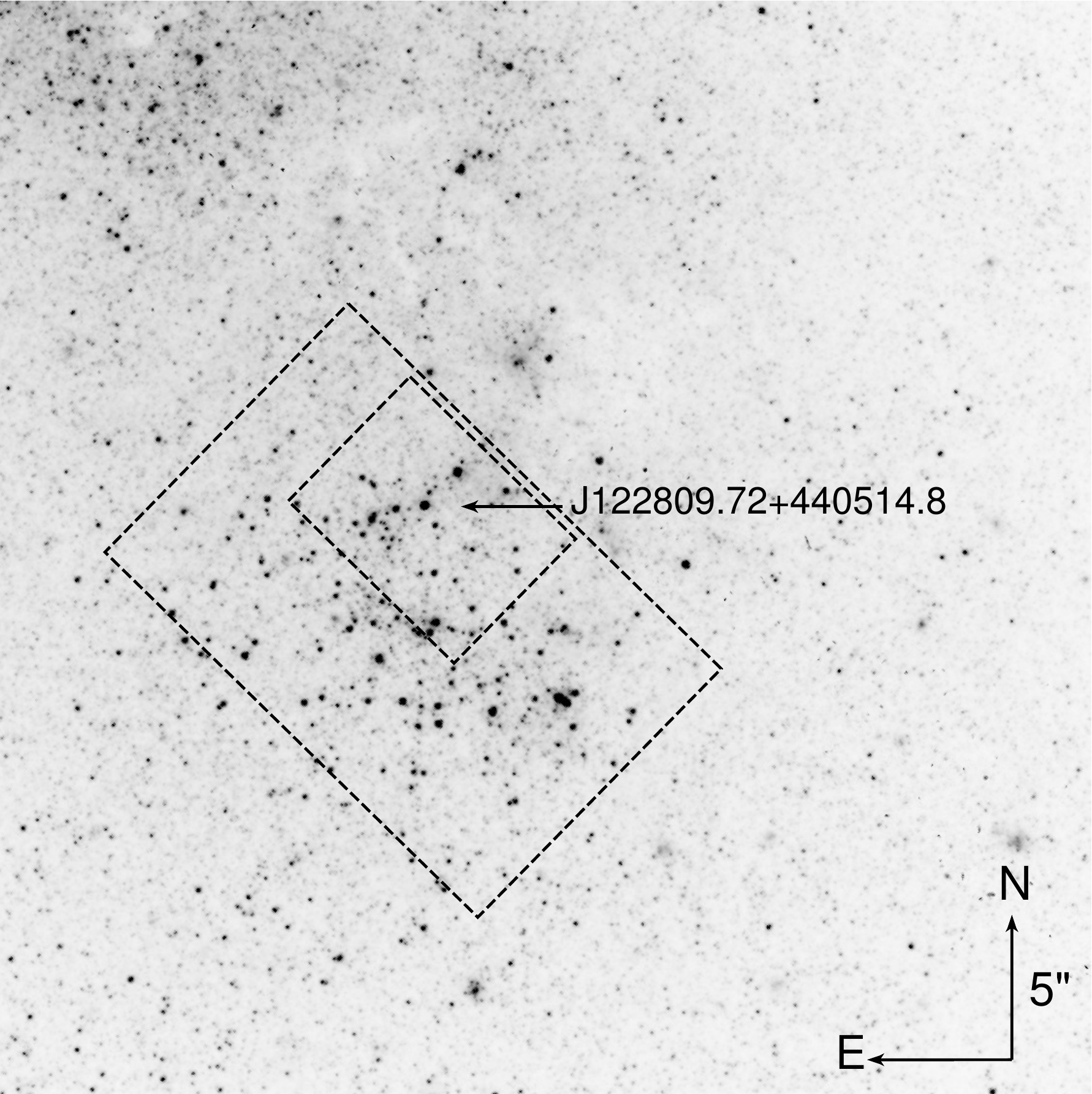}
}
\caption{\textit{HST}/ACS/WFC image in F814W band of J122810.94+440540.6 and J122811.70+440550.9 (left), J122817.83+440630.8 (middle), J122809.72+440514.8 (right). The rectangles represent the star-forming regions that were used to construct the colour-magnitude diagrams. The arrows indicate the objects under study. }
\label{Fig10} 
\end{figure*}

All the stars we study are located near the stellar associations in which these stars were probably formed. To estimate the age of the stellar environment, we have performed PSF photometry and have constructed «colour-magnitude» diagrams (CMD). For J122810.94+440540.6, J122817.83+440630.8 and J122809.72+440514.8, we have selected two regions: the small one, which corresponds to the size of the nearest stellar groups, and the large one, which covers all star-forming region near the star (Fig.~\ref{Fig10}). In the case of J122811.70+440550.9, we have selected two neighbouring large stellar groups. The choice of different regions is due to the uncertainty of the region where the stars were formed.

%Fig.12
\begin{figure*} 
\center{
\includegraphics[angle=0, width=0.40\linewidth]{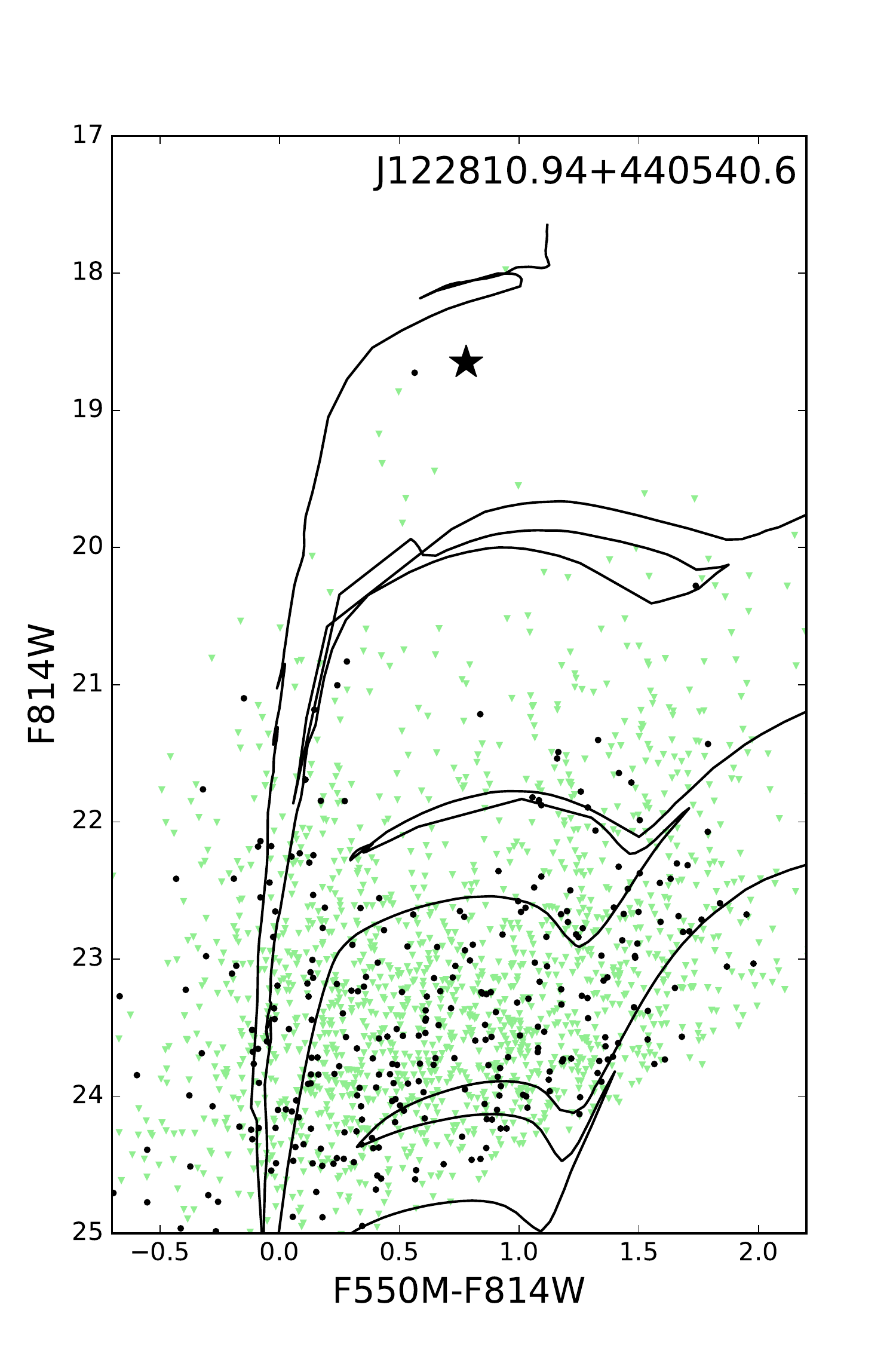}
\includegraphics[angle=0, width=0.40\linewidth]{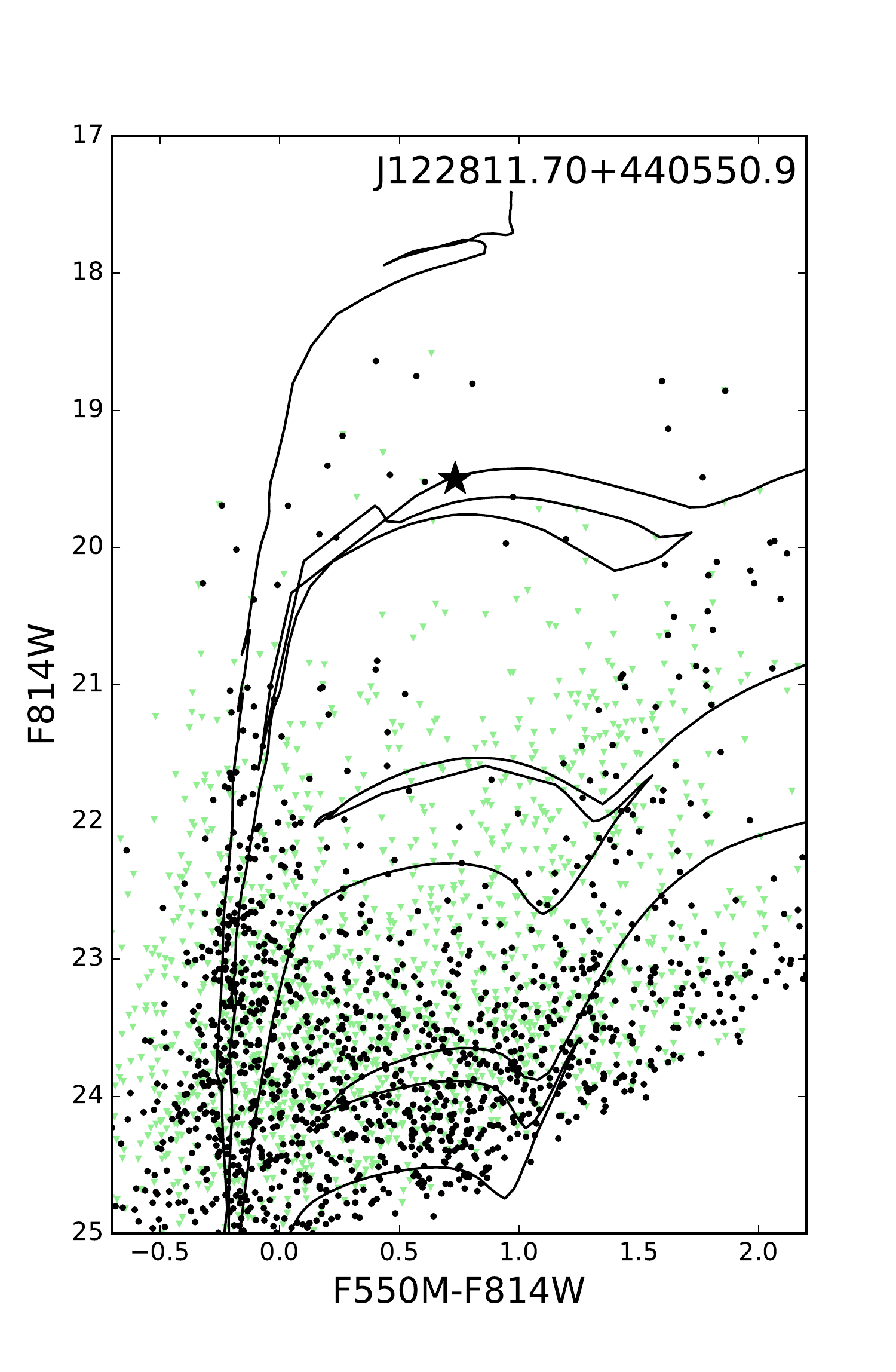}
\includegraphics[angle=0, width=0.40\linewidth]{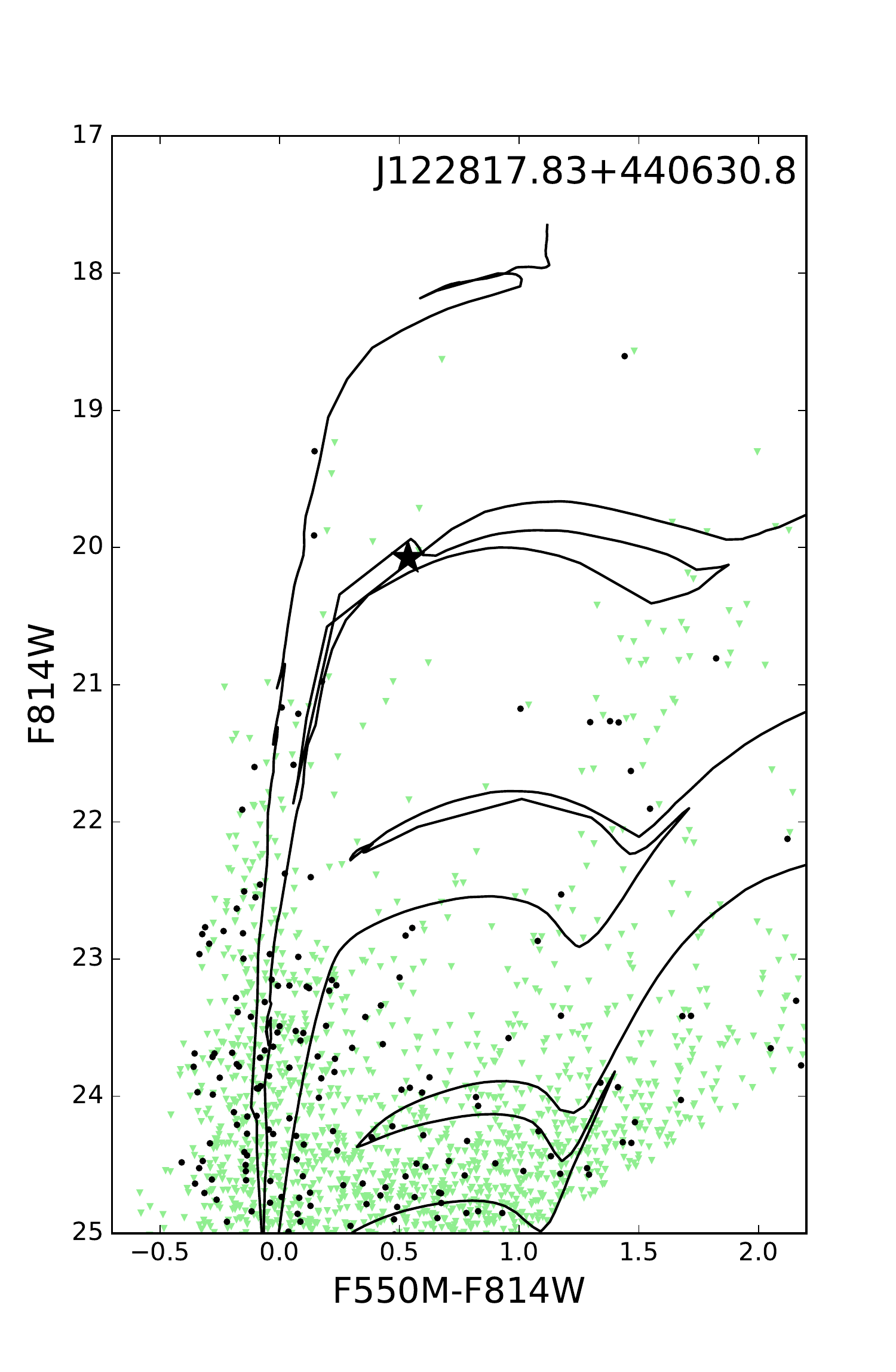}
\includegraphics[angle=0, width=0.40\linewidth]{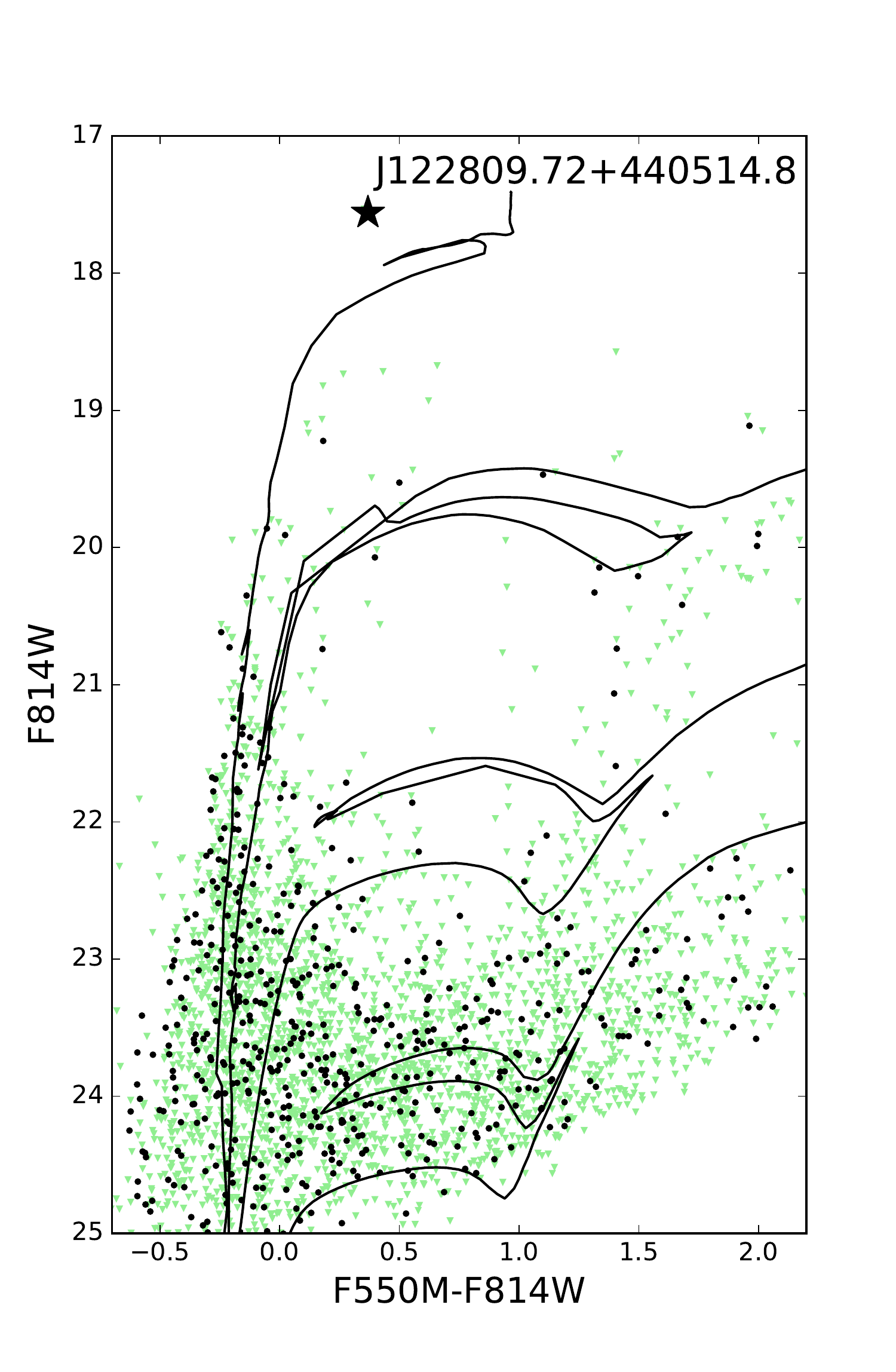}
}
\caption{Color-magnitude diagrams for stellar environments of four stars. The green triangles mark stars of large star-forming regions, the black circles denote stars in small areas around the studied objects, which are shown by black star symbols. Theoretical isochrones of 5, 10, 30 and 100 Myr are shown (from top to bottom).} 
\label{Fig11} 
\end{figure*}

Photometry for the  selected region was performed using archival data obtained in the F550M (2005/11/18) and F814W (2005/11/17) filters of the HST/ACS/WFC using the \textsc{dolphot} package \citep{Dolphin2016}. Fig.~\ref{Fig11} shows colour-magnitude diagrams for the studied associations. The theoretical isochrones\footnote{obtained from http://stev.oapd.inaf.it/cgi-bin/cmd} from \citet{Marigo2017} for the metallicity $Z = 0.5Z_{\odot}$ and the positions of the studied objects are also plotted on the diagram. When calculating the tracks, the canonical two-part-power law corrected for unresolved binaries was chosen as the initial mass function \citep{Kroupa2001, Kroupa2002}. The isochrones were reddened, and extinction was varied from the Galactic value to the maximum value that we measured from the spectral data. The optimal values of reddening were determined individually for each region, by a comparison of the reddened isochrone with the observed main sequence of the stars from selected regions. As a result, we obtained reddening estimates $A_V \approx 0.5^m $ for the stellar environment of J122810.94+440540.6 and J122817.83+440630.8, which is quite consistent with the average internal reddening in NGC 4449 \citep{Hill1998}, and $A_V \approx 0.1^m $ for stellar groups near the J122811.70+440550.9 and J122809.72+440514.8, which turned out to be below the average internal reddening.
 
The positions of stars in the selected regions on the CMD correspond either to continuous star formation within the studied regions for at least the last 100 Myr, or to several episodes of star formation during the same period of time. At the same time, the age of the youngest stars is approximately 5-10 Myr. We suppose that they (like the objects under study) were formed in one of the last bursts of star formation.
Moreover, we did not find a difference between the positions of the stars from the small and large regions on the CMD. More accurate age estimates require information on the values of many parameters such as rotation speed of stars, local metallicity (which may vary even inside one region), initial mass function, etc.
Another factors that can lead to both underestimation and overestimation of the age of the surrounding population are related with the relatively small number of stars in the studied associations. Thus the upper part of the diagram is poorly populated. Therefore, we can overestimate the age of the last starburst, as in the case of J122809.72+440514.8, which is located well above the brightest stars in the diagram. At the same time, there may be unresolved groups of stars among bright stars, which can lead to underestimation of the age of the youngest objects.

\subsection{Masses and luminosities}

\begin{figure}
 \centering
 \includegraphics[angle=0, scale=0.4]{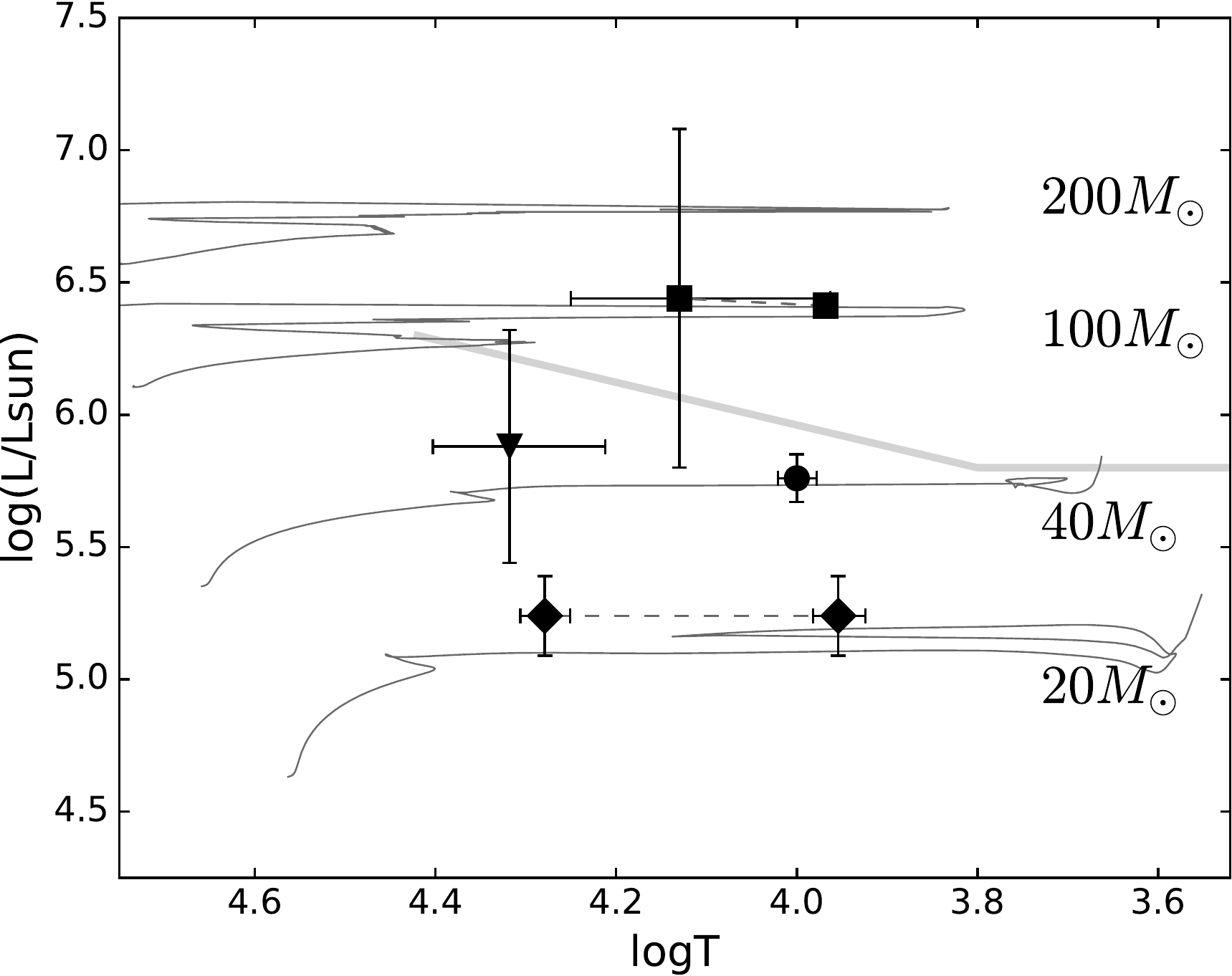}
 \caption{Temperature-luminosity diagram with evolutionary tracks of massive stars for the metallicity of $Z=0.5Z_{\odot}$ \citep{Tang14}. The positions of objects are shown by circle (J122810.94+440540.6), triangle (J122811.70+440550.9), squares (J122809.72+440514.8) and diamonds (J122817.83+440630.8) respectively. Thick grey line indicates the Humphreys-Davidson limit. We also have shown transitions of J122817.83+440630.8 and J122809.72+440514.8 between states by thin grey dash lines.}
    \label{Fig12}
\end{figure}

To estimate the mass of studied objects, we have constructed a temperature-luminosity diagram (Fig.~\ref{Fig12}) and have compared the positions of objects with evolutionary tracks of massive stars of different initial masses \citep{Tang14}. We assumed the metallicity value $Z\approx0.5Z_{\odot}$ \citep{Annibali2017} when choosing evolutionary tracks of stars. An estimates of the photosphere temperatures and luminosities of our stars (in the case of J122809.72+440514.8 for the hot state) are obtained from the SED fitting. The temperature and luminosity of the cold state of J122809.72+440514.8 were taken from the results of modeling with the CMFGEN code. 

Object J122817.83+440630 has an initial mass of about 20 $M\odot$. Stars with such masses can pass the red supergiant stage during their evolution \citep{Humphreys16}. J122809.72+440514.8 has the largest initial mass (more than 100 $M\odot$). The mass loss rate characteristic for stars of such large masses does not allow them to become red supergiants \citep{Humphreys1979} and they evolve into WR stars. The mass values of J122810.94+440540.6 and J122811.70+440550.9 are between the indicated extreme values. Thus, the masses of all four stars are quite typical of LBV stars or massive supergiants.

\section{Conclusions}

In this work, we have studied four massive (from 20 to $\lesssim100\,M\odot$) stars in the NGC\,4449 galaxy. Two stars have shown strong photometric variability for several years, which allows us to classify them as LBVs. Modeling the cold state ($T_{\text{eff}}=9300\,$К) of the brightest one using the CMFGEN code showed a significantly high mass loss rate $\dot{M} = 5.2\times10^{-3}\,M_{\odot}\,yr^{-1}$. This star is a close analogue of $\eta$ Car in its state.

The classification of the remaining two stars is complicated by their small photometric variability and low signal-to-noise ratio in their spectra, which do not allow us to reliably confirm or exclude their own emission in such characteristic lines as [\ion{O}{i}]6300 and/or [\ion{Ca}{ii}]. The presence of a significant contribution of both f-f and f-b radiation to the SED of one of them indicates the presence of an ionized shell, characteristic of B[e]-supergiants. However, a more precise classification of both stars requires further observations.

\section*{Acknowledgements}
This research was supported by the Russian Foundation for Basic Research 19-52-18007. Observations with the SAO RAS telescopes are supported by the Ministry of Science and Higher Education of the Russian Federation (including agreement No05.619.21.0016, project ID RFMEFI61919X0016). The renovation of telescope equipment is currently provided within the national project ''Science''. The modelling spectra was performed as part of the government contract of the SAO RAS approved by the Ministry of Science and Higher Education of the Russian Federation. The research made use  of equipment purchased with funds from the Program of Development of M.\,V.\,Lomonosov Moscow State University. 

\section*{Data Availability}
The data underlying this article will be shared on reasonable request to the corresponding author. 

%%%%%%%%%%%%%%%%%%%%%%%%%%%%%%%%%%%%%%%%%%%%%%%%%%



\bibliographystyle{mnras} \bibliography{bibtexbase.bib}

%\begin{thebibliography}{99}
%\end{thebibliography}

%%%%%%%%%%%%%%%%%%%%%%%%%%%%%%%%%%%%%%%%%%%%%%%%%%

%%%%%%%%%%%%%%%%% APPENDICES %%%%%%%%%%%%%%%%%%%%%

% \appendix

% \section{Some extra material}

% If you want to present additional material which would interrupt the flow of the main paper,
% it can be placed in an Appendix which appears after the list of references.

%%%%%%%%%%%%%%%%%%%%%%%%%%%%%%%%%%%%%%%%%%%%%%%%%%

% Don't change these lines
\bsp	\label{lastpage} \end{document}